\documentclass{aa}
\usepackage{graphicx}
\def\ssim{\setbox0=\hbox{$\propto$}%
\setbox1=\hbox{$<$}\dimen0=\ht1%
\advance\dimen0by-1.2pt\,\lower.6\dimen0%
\copy0\kern-\wd0\raise.4\dimen0\copy1 \,}

\def\gsim{\setbox0=\hbox{$\propto$}%
\setbox1=\hbox{$>$}\dimen0=\ht1%
\advance\dimen0by-1.2pt\,\lower.6\dimen0%
\copy0\kern-\wd0\raise.4\dimen0\copy1\,}

\def\lambdab{\lambda\mkern-9mu\lower1.2pt\hbox{$\mathchar'26$}}%

\begin{document}
   \title{Stellar evolution with rotation X: }

\subtitle{Wolf-Rayet star populations at solar metallicity}

 \author{G. Meynet, A. Maeder}

     \institute{Geneva Observatory CH--1290 Sauverny, Switzerland\\
              email:  Georges.Meynet@obs.unige.ch\\
              email: Andre.Maeder@obs.unige.ch }

   \date{Received 5 February 2003/ Accepted 6 March 2003}

\abstract{We examine the  properties of Wolf--Rayet (WR) stars predicted by models
of rotating stars taking account of the new mass loss rates for O--type stars
and WR stars (Vink et al. \cite{Vink00}, \cite{Vink01}; Nugis \& Lamers \cite{NuLa00})
and of the wind anisotropies induced by rotation. We find that the rotation velocities $v$
of  WR stars are modest, i.e. about 50 km s$^{-1}$, not very dependant on the initial $v$
and masses. For the most massive stars, the evolution of $v$ is very strongly influenced
by the values of the mass loss rates; below $\sim$12 M$_\odot$  the evolution of rotation
during the MS phase and later phases is dominated by the internal coupling. Massive stars
with extreme rotation may skip the LBV phase.

Models having a typical $v$ for the O--type stars have WR lifetimes on the average two times longer
than for non--rotating models. The increase of the WR lifetimes is mainly due to that of
the H--rich eWNL phase. Rotation allows a transition WN/WC phase to be present for
initial masses lower than 60 M$_\odot$. The  durations of the other WR subphases are
less affected by rotation. The mass threshold for forming WR stars is lowered from
37 to 22 M$_\odot$ for typical rotation. The comparisons of the predicted number ratios  
WR/O, WN/WC  and of the number of transition WN/WC stars show very good agreement with models with
rotation, while this is not the case for models with the present--day mass loss rates 
and no rotation. 
As to the chemical abundances in WR stars, rotation brings only very small changes 
for WN stars, since they have equilibrium CNO values. However, WC stars with rotation
have on average lower C/He and O/He ratios. The luminosity distribution of WC stars
is also influenced by rotation.
 }

   \maketitle
%

\section{Introduction}

Wolf--Rayet stars play an important role in Astrophysics, as signatures
of star formation in galaxies and starbursts, as injectors of chemical elements and of the 
radioactive isotope $^{26}$Al, as  sources of kinetic energy into the interstellar medium and 
as progenitors of supernovae and maybe as progenitors of $\gamma$--ray bursts.
As rotation affects all outputs of stellar models (Meynet \& Maeder \cite{MMV}),
the main purpose of the present paper is to examine some 
consequences of rotation on the properties of Wolf--Rayet (WR) stars.

Let us recall
some difficulties faced by non--rotating stellar models of WR stars.
A good agreement between 
the predictions of the stellar models for the WR/O number ratios and the observed 
ones at different metallicities in regions of constant star formation was achieved 
provided the mass loss rates were enhanced by about a factor of two during the MS
and WNL phases (Meynet \& al. \cite{Mey94}). This solution, which at that time appeared 
reasonable in view of the uncertainties pertaining to the mass loss rates, is no longer
applicable at present. Indeed, the estimates of the mass loss rates during the WR phase are
reduced by a factor of 2 to 3, when account is taken of the clumping effects
in the wind (Nugis and Lamers \cite{NuLa00}; Hamann and Koesterke \cite{Ha99}).
Also, the mass loss rates for O--type stars have been substantially revised 
(and in general reduced) by the new results of 
Vink et al. (\cite{Vink00}, \cite{Vink01}), who also account for the occurrence of bi--stability
limits which change the wind properties and mass loss rates.
In this new context, it is quite clear that with these new mass loss rates
the predicted numbers of WR stars by standard non--rotating models would be much
too low with respect to the observations.

A second difficulty of the standard models with mass loss was
the observed number of transition WN/WC stars. These stars simultaneously show some 
nitrogen characteristics of WN stars and some carbon of the further WC stage.
The observed frequency of WN/WC stars among WR stars is around 4.4 \% (van der Hucht \cite{Hu01}), while
the frequency predicted by the standard models without extra mixing processes 
are lower by 1--2 orders of magnitude (Maeder \& Meynet \cite{MaeMey94}).
A third difficulty of the standard models as far as WR stars were concerned was that
there were relatively too many WC stars with respect to WN stars predicted (see the review
by Massey \cite{Mas03}). These difficulties are the signs that some process is missing
in these models.

We explore here the effects of rotation on the WR stars and their formation.
Rotation  already improved
many predictions of the stellar models, bringing better agreement 
with the observations. In particular,  star models with rotation have at
last enabled us (Maeder \& Meynet \cite{MMVII})
to solve the long--standing problem of the large number of 
observed red supergiants at low metallicities $Z$ in the SMC, while current models
with any kind of reasonable mass loss usually predicted almost no red
supergiants at low $Z$.
Also, in addition to several 
other points, the new models
also show significant He-- and N--enhancements in O--type stars on the
MS (Heger and Langer \cite{He00}; Meynet \& Maeder \cite{MMV}), as well as in supergiants at various metallicities, as required by observations.

Sect. 2 summarizes the physics of these models. The evolution of the inner and surface 
rotation is discussed in Sect. 3.  The evolutionary tracks are shown in Sect. 4.
The effects of rotation on the WR star formation process and on the WR lifetimes
are discussed in Sect. 5. Comparisons with the observed WR populations are
performed in Sect. 6. Finally the predicted abundances in WR stars are discussed in Sect. 7.

\section{Physics of the models}

The present grid of models at solar metallicity is based in general 
on the same physical assumptions
as in the grid by Meynet \& Maeder (\cite{MMV}, paper V). However, 
in addition we include here several improvements that have appeared since.
These are:

\begin{itemize}

\item As reference mass loss rates in the case of no rotation, 
we use the recent data by Vink et al. (\cite{Vink00}; \cite{Vink01}).
A bi--stability limit intervenes at  T$_{\mathrm{eff}}= 25000$ K
and produces some steep changes of the mass
loss rates during MS evolution. For the domain not covered by these authors
we use the results by de Jager et al. (\cite{Ja88}).
Since the empirical values
for the mass loss rates 
employed for non--rotating stars are based on 
stars covering the whole range of rotational velocities, we must apply a reduction factor to the empirical rates to make
them correspond to the non--rotating case. The same reduction factor as
in Maeder \& Meynet (\cite{MMVII}) was used here.

\item During the Wolf--Rayet phase we apply
the mass loss rates proposed by Nugis and Lamers (\cite{NuLa00}). These mass loss rates,
which account for the clumping effects in the winds,  
are smaller by a factor 2--3 than the mass loss rates used in our previous 
non--rotating stellar grids (Meynet et al. \cite{Mey94}).

\item The significant increases of the mass loss rates $\dot{M}$ with  
rotation are taken into account as explained in Maeder \& Meynet (\cite{MMVI}). This treatment,
in contrast to the one usually used (Friend and Abbott \cite{Fr86}), accounts for 
the non--uniform brightness of rotating stars
due to the von Zeipel theorem and for the fact that the Eddington
factor is also a function of the rotation velocity. 
Due to the non--uniform brightness of a rotating star, the stellar winds
are anisotropic. The rotation--induced anisotropy of the stellar winds
is also accounted for during the Main Sequence (MS) phase according 
to the prescription of Maeder (\cite{Ma02}).
For O--type stars, the polar winds are enhanced and this means that these
stars lose a lot of mass, but a moderate amount of angular momentum. 
For stars cooler than T$_{\mathrm{eff}}= 25000$ K, the equatorial ejection
becomes significant and leads to a larger ejection of angular
momentum.  After the MS phase, 
the ratios of the surface angular velocity to the break--up velocity
are in general much too low for the anisotropies to be important.  Interestingly enough,
in cases of extreme mass loss such as in WR stars, the timescale for the transmission
of the extra torque applied at the stellar surface by the anisotropies
is longer than the timescale for the mass removal of the considered layers,
  so that the effects of the anisotropies
of the stellar winds  on the internal rotation are rather limited 
(Maeder \cite{Ma02}).

\item In paper V, we did not account for the effects of
horizontal turbulence on the shears (Talon \& Zahn \cite{TaZa97}). We account for it in the
present work in the same manner as we did in paper VII (Maeder \& Meynet
\cite{MMVII}). Let us recall that the horizontal turbulence, expressed by
a diffusion coefficient $D_{\mathrm{h}}$, tends to reduce
 the shear mixing in regions of steep 
$\mu$--gradients, making the diffusion coefficient $D_{\rm shear}$ proportional
to $D_{\mathrm{h}}$ rather than to $K$, the thermal diffusivity.
 The difference can be large and it leads to smaller 
surface enhancements in the products of CNO burning.  In regions of low 
$\mu$--gradients, the horizontal turbulence moderately increases the shear diffusion
coefficient, making it proportional to  $D_{\mathrm{h}} + K$ rather than to $K$ only, with 
little consequence for the surface enrichments.

\item A moderate overshooting is included in the present
rotating and non--rotating models. The radii of the convective cores are increased
with respect to the values obtained by the Schwarzschild criterion by a quantity
equal to 0.1 H$_{\rm p}$, where H$_{\rm p}$ is the pressure scale height evaluated at the Schwarzschild
boundary. 

\item As in paper VIII (Meynet \& Maeder \cite{MMVIII}), the opacities are
from Iglesias \& Rogers (\cite{IR96}), complemented at low
temperatures with the molecular opacities of Alexander
(http://web.physics.twsu.edu/alex/wwwdra.htm). The nuclear
reaction rates are also the same as in paper VIII and are
based on the new NACRE data basis (Angulo et al. \cite{Ang99}).

\item The initial composition is changed with respect to paper V:
the mass fraction of hydrogen is X = 0.705, the mass fraction of helium Y = 0.275.
For the heavy elements, 
we adopt the same mixture as the one
used to compute the opacity tables for solar composition.
\end{itemize}

The effect of rotation on the transport of the chemical species and the
angular momentum are accounted for as in our papers VII and VIII.
All the models were computed 
up to the end of the helium--burning phase.

\section{Evolution of the rotation}

\subsection{Global view: the effects of coupling and  mass loss}

The evolution of the rotation velocities at the stellar surface depends mainly 
on 2 factors, the internal coupling and the mass loss.

\begin{figure}[t]
  \resizebox{\hsize}{!}{\includegraphics{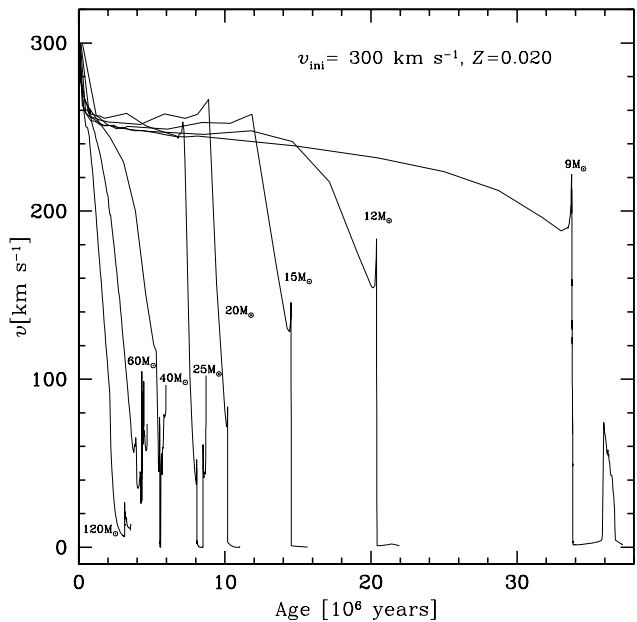}}
  \caption{Evolution of the rotational velocities for star models 
  of different initial masses between 120 and 9 M$_{\odot}$ with account taken
  of anisotropic mass loss during the MS phase. The rapid decrease near the end 
  of the MS evolution is due to the bi--stability limit in the mass loss rates.
   The location  of the end of the MS phase corresponds to the vertical decrease of the velocity.}
  \label{veq}
\end{figure}

1.-- The  coupling mechanisms  transport angular momentum
in the stellar interiors. The extreme case of strong coupling is
the classical case of solid body rotation. In this case when mass loss is small, the star  
reaches the critical velocity during the MS phase more or less quickly depending on the
initial rotation as shown by Sackman \& Anand (\cite{Sa70}; see also Langer \cite{La97}).
In that case, coupling ensures that $\Omega/\Omega_c$ increases when the radius of the star increases.
Let us recall that for $\Omega/\Omega_c$ to increase when the radius of the star increases,
its suffices that $\Omega$ decreases less steeply with the radius than $\Omega_c$ ($\propto R^{-3/2}$), or
expressed in another way, if
$\Omega \propto R^{-\alpha}$ then $\Omega/\Omega_c$ increases with the radius when $\alpha < 3/2$.
In the case of no coupling, i.e. of local conservation of the angular momentum,
rotation becomes more and more subcritical 
($\alpha=2$). In the present models, the
situation  is  intermediate, with a moderate coupling 
due mainly to meridional circulation, which is more efficient 
(Meynet \& Maeder \cite{MMV}) than shear transport,
as far as transport of angular momentum is concerned.  

2.-- For a given degree of coupling, {\emph{the mass loss rates play a most critical role 
in the evolution of the surface rotation}}. As shown by the comparison of the
models at $Z=0.02$ and $Z=0.004$ (Maeder \& Meynet \cite{MMVII}), for masses greater
than 20 M$_{\odot}$ the models with solar composition have velocities that decrease
rather rapidly, while at $Z=0.004$ the velocities go up. Thus, for the most massive stars 
with moderate or strong coupling, the mass loss rates are the main factor influencing 
the evolution of rotation.

\begin{figure}[t]
  \resizebox{\hsize}{!}{\includegraphics{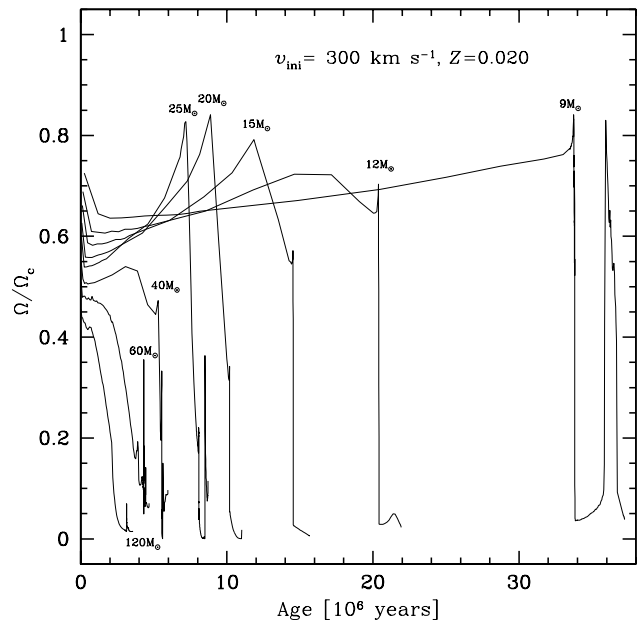}}
  \caption{Evolution of the fraction 
  $\frac{\Omega}{\Omega_{\mathrm{c}}}$ of the angular velocity to 
  the critical angular velocity at the surface   of  star models 
  of different initial masses between 120 and 9 M$_{\odot}$ with account 
  taken of anisotropic mass loss during the MS phase.}
  \label{omc}
\end{figure}

Below a mass of about 12 M$_{\odot}$, the mass loss rates are smaller and the internal 
coupling plays the main role in the evolution of the rotational velocities. This
provides an interesting possibility of tests on the internal coupling by studying
the differences in rotational velocities for stars at different distances of the 
ZAMS. In particular, such a study could allow us  to test the role
of magnetic coupling in radiative envelopes, which is now a major open question
in stellar rotation studies (Spruit \cite{spruit02}).

Figs.~\ref{veq} and \ref{omc} show the evolution of the rotational velocities and of the fraction 
 $\frac{\Omega}{\Omega_{\mathrm{c}}}$ of the angular velocity to 
 the critical angular velocity at the surface   of  star models 
 of different initial masses between 9 and 120 M$_{\odot}$ with account taken
 of anisotropic mass loss during the MS phase. Globally these figures are
 rather similar to those of previous models (Meynet \& Maeder \cite{MMV}), except for the fact 
 that the mass loss rates by Vink et al. (\cite{Vink00}, \cite{Vink01}) are used here.
 These rates  meet a bi--stability limit at $\log T_{\mathrm{eff}}=4.40$.
 Below this value there is a sudden increase of the mass loss rates, which makes
 the rotation velocities rapidly decrease, as for the most massive stars.
This is clearly visible for the models below 40 M$_{\odot}$ in Figs.~\ref{veq} and \ref{omc}. 
Thus, for these models with a rapidly decreasing velocity of rotation, 
we again stress that any comparison between observed and predicted rotation for these
domains of mass is really much more a test of the mass loss rates than a test on
the internal coupling and evolution of rotation. We may mention that for an initial velocity $v_{\mathrm{ini}} 
\leq 300$ km s$^{-1}$, the account taken of the anisotropic wind does not play a major
role (see Sect. 4).

\begin{figure}[t]
  \resizebox{\hsize}{!}{\includegraphics{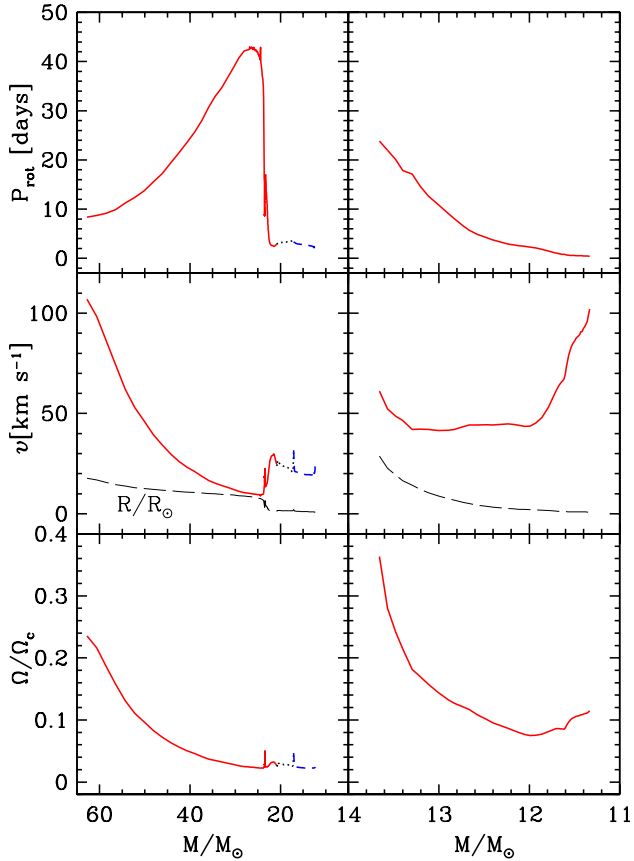}}
  \caption{Evolution as a function of the actual mass 
  of the rotation period, the rotational velocity
  and the ratio of the angular velocity to the critical value during
  the WR stage of massive stars. The continuous line corresponds
to the so--called eWNL phase (see text), the dotted and short--dashed lines show the evolution
during the eWNE and WC phases respectively. The long--dashed lines
in the panels for the velocities show the evolution of the polar radius,
(in this case the unity on the ordinates are $R/R_{\odot}$).
{\it Left}: the WR phase of a star with an initial
  mass of 85 M$_{\odot}$ with $v_{\mathrm{ini}}= 300$ km s$^{-1}$. 
{\it Right}: the same for an initial mass of 25 M$_{\odot}$ with $v_{\mathrm{ini}}= 300$ km s$^{-1}$.}
  \label{vrotwr1}
\end{figure}
Thus, the uncertainties
on the mass loss rates and bi--stability limits
prevent us from making significant tests on the internal physics and
coupling in the massive stars for models with initial masses above about
12 M$_{\odot}$. Mass loss uncertainties dominate the problem. From the rather large differences 
between the values of the mass loss rates published over recent years, we know that
these uncertainties are still great. Fortunately, below 12 M$_{\odot}$ 
the comparisons may be meaningful and may reveal whether our models have the right 
kind of internal coupling.

\subsection{Evolution of the rotational velocities during the WR stages}

 Fig.~\ref{vrotwr1} (left) shows the evolution of the rotation period $P$, of the
rotation velocities $v$ and of the fraction 
 $\frac{\Omega}{\Omega_{\mathrm{c}}}$ of the angular velocity to 
 the critical angular velocity at the surface 
during the WR stages of a star model with an initial mass of
85 M$_{\odot}$ with $v_{\mathrm{ini}}= 300$ km s$^{-1}$
typical of a star which has both a WN and WC phase. 
During the WN stage of the 85 M$_{\odot}$ model, 
until the actual mass of the star reaches a value near 23 M$_\odot$,
the angular velocity decreases by a factor of $\sim 5$, 
correspondingly the period
$P = 2 \pi / \Omega$ grows by the same factor and the rotation velocity $v= \Omega R_{\mathrm{e}}$
decreases much more since the equatorial radius $R_{\mathrm{e}}$ also decreases during this phase
(by a factor of $\sim 2$). The fraction $\frac{\Omega}{\Omega_{\mathrm{c}}}\simeq
\frac{\Omega \; {R_{\mathrm{e}}^{\frac{3}{2}}}}   {(G \; M)^{\frac{1}{2}}   }$
also decreases very much, because the sensitivity to the decrease of the radius is
larger than that to the mass (which decreases by a factor of  $\sim 2.7$).

We adopt here the definition of the WN phases given by Foellmi et al.
(\cite{Fo03}), putting a prefix  ``e'' when the class is assigned by composition properties of
the evolutionary models.
Thus, eWNE means WN stars without hydrogen, while eWNL means WN stars with hydrogen,
(see also Sect 5 below).
When the  H--content goes to zero (i.e. just before the star enters the eWNE
stage), we see a dramatic
increase of $\Omega$ (or decrease of $P$), due to the strong contraction of the surface layer
because the opacity of the surface layers becomes much smaller. 
This transition is fast, thus the transfer of angular momentum
by meridional circulation is limited and the evolution of the rotation at the surface
is dominated by the local conservation of angular momentum, which explains 
the rapid and large increase of $\Omega$. However, we see that 
despite the strong decrease of $P$, the velocity $v$  and the ratio
$\frac{\Omega}{\Omega_{\mathrm{c}}}$ do not change
very much, since the radius is much smaller in the eWNE stage.
When the star becomes a WC star with an actual mass of $\sim$ 17.3 M$_\odot$,
there 
is very little change in the velocity, since  rotation is already very low
and the radius no longer decreases significantly.

Fig.~\ref{vrotwr1} (right) shows the same for a star 
of 25  M$_{\odot}$ with $v_{\mathrm{ini}}= 300$ km s$^{-1}$, which does not enter the 
eWNE and WC stages. We see that $P$ decreases (or $\Omega$ increases) during 
the WN stage. This is because the star model moves from 
$\log T_{\mathrm{eff}}$ 4.29 to 4.70. The decrease of the radius
makes the star rotate faster, slightly overwhelming the mass loss effects 
from 13.65 to  11.33 M$_{\odot}$. The increase of $\Omega$ and the decrease
of  $R$ nearly compensate for each other so that $v$ does not vary much
during most of the WN phase, except at the very end. There the increase of $\Omega$
is more rapid than the decrease of the radius and an increase in the velocity
is obtained. The reason for this behaviour is that in this model the
 stellar winds uncovers the layers, where the $\Omega$--gradient is steep, and
 we start seeing layers which were deep inside and rotating rapidly.

\begin{figure}[t]
  \resizebox{\hsize}{!}{\includegraphics{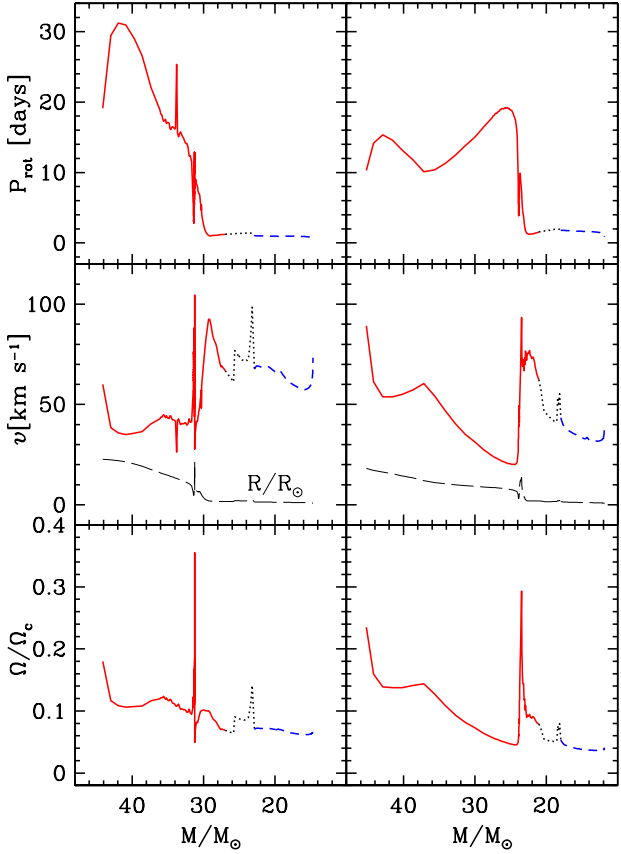}}
  \caption{Same as Fig.~\ref{vrotwr1} for 60 M$_\odot$ stellar models.
{\it Left}: the WR phase of a star with an initial
  mass of 60 M$_{\odot}$ and $v_{\mathrm{ini}}= 300$ km s$^{-1}$.
{\it Right}: for an initial mass of 60 M$_{\odot}$ with $v_{\mathrm{ini}}= 500$ km s$^{-1}$.
   In the e--version, the red color corresponds to the phase where H is 
   present (eWNL), the black to the phase of an He--star (eWNE) and the blue to
   the WC phase.}
  \label{vrotwr2}
\end{figure}

Globally, we note  that the rotational velocities of WR
stars are modest in the 2 cases examined above, despite the very different masses and 
evolution stages. 
This is firstly the result of the strong mass loss necessary to produce the WR stars
and also of the mass loss during the WR stages. 
We note however that, for a given initial velocity, the velocities obtained at the end of the He--burning phase
for the WR stars decrease when the initial mass increases (see Table~\ref{tbl-1}).
This results from the simple fact that the higher  the initial mass, the larger  the quantity
of mass (and thus of angular momentum) removed by the stellar winds.

\begin{figure}[ht]
  \resizebox{\hsize}{!}{\includegraphics{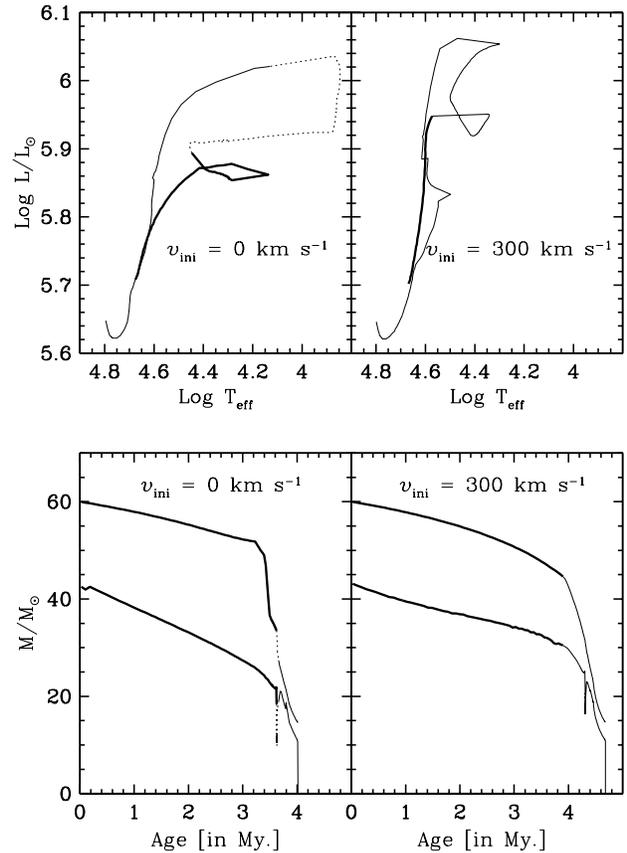}}
  \caption{The two upper panels show the evolutionary tracks of a non--rotating and a rotating 60 M$_\odot$ stellar model.
The bold part of the track corresponds to the O--type star Main--Sequence phase; the dotted part shows the intermediate phase,
when present, between the O--type star phase and the WR phase; the thin continuous line is the track during
the WR phase. The two lower panels show the evolution as a function of time of the total mass and of the masses of the
convective cores during the H-- and the He--burning phases.
}
  \label{hrkip}
\end{figure}

Figs.~\ref{vrotwr2} shows the evolution of rotation
properties for a 60 M$_{\odot}$ model with $v_{\mathrm{ini}}= 300$ km s$^{-1}$ (left)
and with $v_{\mathrm{ini}}= 500$ km s$^{-1}$ (right). On the whole, these two cases are not 
very different from the case of the 85 M$_{\odot}$ model shown above. In 
Fig.~\ref{vrotwr2} (left), we firstly see an increase of  $P$ and a corresponding decrease of 
$v$  and of $\frac{\Omega}{\Omega_{\mathrm{c}}}$; this decrease of the rotation is due,
as before, to the high mass loss rates. Then, 
when the actual mass is about 38 M$_\odot$,
$\Omega$ increases again due a decrease of the stellar radius as the star is
moving to the left in the HR diagram (Fig.~\ref{hrkip}).
The velocity $v$ does not change very much since the stellar radius is decreasing.
When the actual mass of the star approaches 30 M$_\odot$,
at an age of $4.307 \cdot 10^6$ yr, 
$X_{\mathrm{s}}= 0.10$ and the outer stellar layers become
more transparent, so that the star brightens rapidly above $\log L/{\rm L}_{\odot}=6.0$ for a short time
until an age of $4.33 \cdot 10^6$ yr. The corresponding variations in radius
produce a sharp oscillation  in $\Omega$ and velocity. Then at an age of  
$4.38 \cdot 10^6$ yr.  until $4.44 \cdot 10^6$ yr., the star 
becomes an He--star (eWNE stage).
The angular velocity remains nearly constant (cf. the evolution of the period)
while the velocity and $\Omega/\Omega_{\mathrm{c}}$ present some more variations
linked to changes in the radius.
Then, the star enters the WC stage, the mass loss rates are so large that despite 
the decrease of the radius, the rotation remains small. As was the case before,
the presence of steep gradients of $\Omega$ at the surface, may result in some cases in an
increase of $\Omega$ when the star is peeled off by the stellar winds.

The case of  60 M$_{\odot}$ with $v_{\mathrm{ini}}= 500$ km s$^{-1}$ (right) contains 
features similar to the previous cases. The 
beginning of the curve, with the first bump of $P$, behaves like the case of
$v_{\mathrm{ini}}= 300$ km s$^{-1}$ of 60 M$_{\odot}$. After the  dip of
$P$ at an age of $4.08 \cdot 10^6$ yr (actual mass of $\sim$ 37 M$_\odot$), the further evolution of $P$
is like the case of an 85 M$_{\odot}$ star, marked by a large decrease of
the angular velocity $\Omega$. Later, when the star becomes an He--star at an age 
$4.50 \cdot 10^6$ yr (actual mass $\sim$ 24 M$_\odot$), we
see a sharp increase 
in $\Omega$,  in the velocity and in the ratio $\Omega/\Omega_{\mathrm{c}}$.
During the WC phase, the velocity remains between 30 and 40 km s$^{-1}$,
surprisingly, this is
a factor $\sim$ 2 below the velocity reached during this phase by the 300 km s$^{-1}$ stellar
model. The lower value obtained for the initially faster rotating star
results from the greater quantity of mass and thus of angular momentum lost during the 
previous evolutionary stages.

\begin{figure*}[ht]
  \resizebox{\hsize}{!}{\includegraphics[angle=-90]{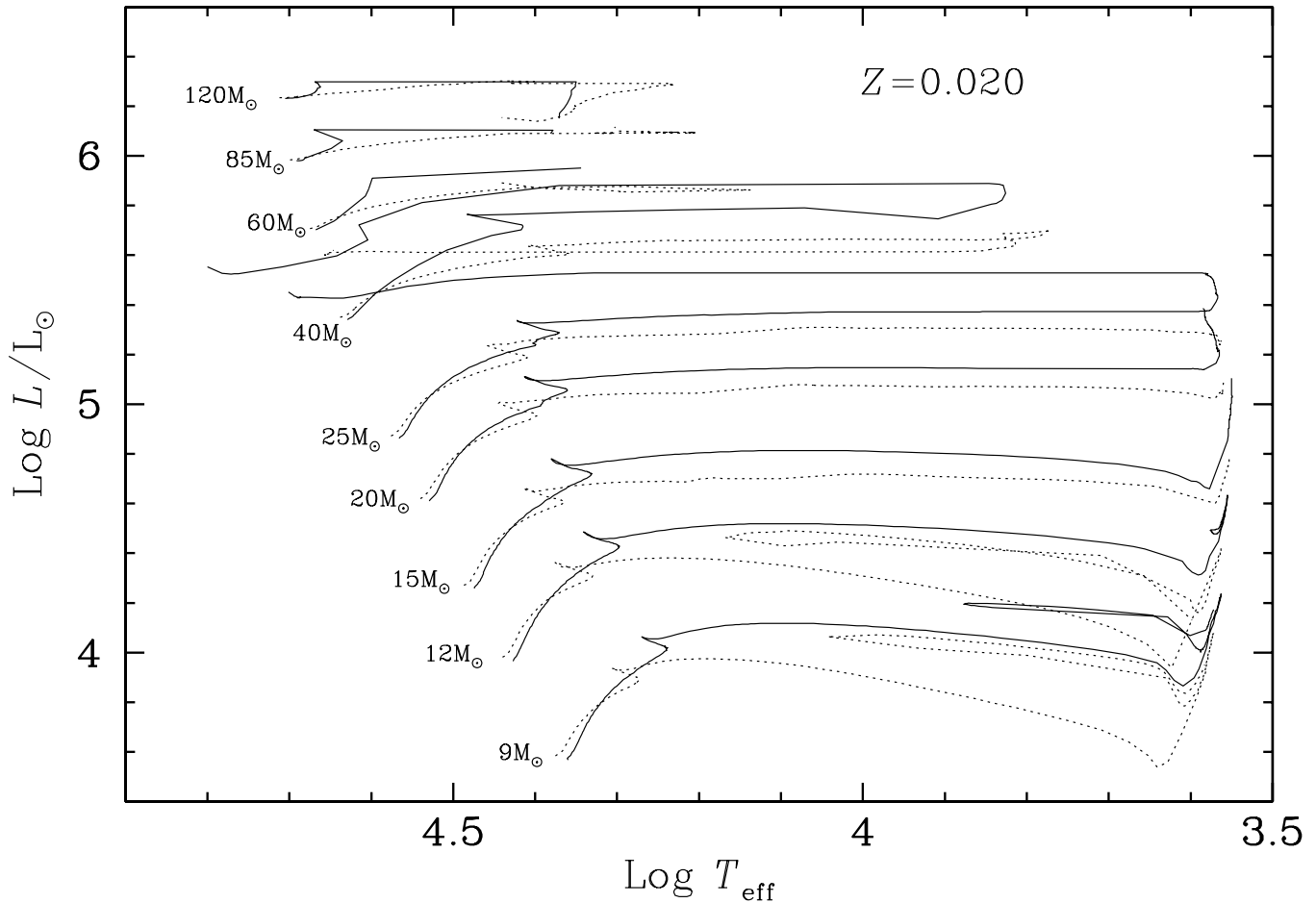}}
  \caption{Evolutionary tracks for non--rotating 
(dotted lines) and rotating (continuous lines) models for a metallicity $Z$ =
0.020. 
The rotating models
have an initial velocity $v_{\rm ini}$ of 300 km s$^{-1}$, which corresponds to an average velocity
during the MS phase of about 180 to 240 km s$^{-1}$ (see Table~\ref{tbl-1}).
}
  \label{hr}
\end{figure*}

The main conclusion of this Section is that {\emph{the predicted surface 
rotation velocities of WR stars are relatively low, of the order of 50 km s$^{-1}$}}.
 Also, in the stellar interiors, rotation is modest with ratios 
$\frac{\Omega_{\mathrm{c}}}{\Omega_{\mathrm{s}}}$ of the angular velocity
at the center to that at the surface which are only of the order of a few 
unities. This is the result of the very strong mass loss 
peeled off the star to make it a WR star. 
It is to be examined whether
the further evolution  and strong central contraction to the pre--supernova stage is
able to accelerate enough to account for the observed rotation periods
of pulsars and maybe also, for rotations fast enough,
for enabling the collapsar model to work (Woosley \cite{Woo03}).
A word of caution is necessary, since there are many paths leading to WR stars
and we cannot exclude that some, possibly rare paths, may lead to WR stars with
very high rotation.

\section{Evolutionary tracks, masses and lifetimes}

Fig.~\ref{hr} presents the ~tracks for non--rotating and rotating evolutionary 
models  for the whole mass range considered.
As a result of the change of the opacities (slightly enhanced) and of the
 initial abundances ($Y$ = 0.275, instead
of 0.30), the non--rotating tracks are shifted to lower effective temperatures
with respect to paper V. The tracks also extend towards lower effective temperatures
during the MS phase.
This is a well known consequence of  overshooting.
Overshooting and core extension by rotation are also
generally   responsible for  shorter
extensions of the blue loops in the He--burning phase. This is what we 
observe in Fig.~\ref{hr}: the extension of the blue loops is further
reduced by rotation in the case of the 9 M$_{\odot}$ model, and the blue loops
are even suppressed in the rotating model of 12 M$_{\odot}$.

\begin{table*}
\caption{Properties of the stellar models at the end of the H--burning phase
and at the end of the He--burning phase. The masses are in solar mass, the velocities in km s$^{-1}$, the lifetimes in million years and the abundances in mass fraction.} \label{tbl-1}
\begin{center}\scriptsize
\begin{tabular}{ccc|ccccccc|cccccc}
\hline
    &         &         &         &         &     &      &      &      &     &      &      &      &         &   &          \\

M & $v_{\rm ini}^1 $ & $\overline{v}$ & \multicolumn{7}{|c|}{End of H--burning} &\multicolumn{6}{|c}{End of He--burning}      \\
    &         &         &         &         &     &      &      &      &     &      &      &      &         &   &          \\
    &  &   & $t_H$ & $t_{\rm O}$ & M & $v$ & Y$_s$ & N/C & N/O &  $t_{He}$ & M & $v$ & Y$_s$ & N/C &  N/O \\
    &         &         &         &         &     &      &      &      &     &      &      &      &         &    &         \\
\hline
    &         &         &         &         &     &      &      &      &     &      &      &      &         &    &         \\
120 &  0 & 0 &  2.738 & 1.970 & 42.662 &   0 & 0.96 & 46.0 & 47.0 &   0.356  & 16.293 &  0  & 0.27 & 0 & 0  \\
    & 300 & 200 &  3.138 & 2.153 & 26.235 &  9.5 & 0.93 & 50.8 & 43.6 &  0.407 & 11.297 & 13.9  & 0.30 & 0    & 0     \\
    &         &         &         &         &     &      &      &      &     &      &      &      &         &     &        \\
 85 &  0 & 0 &  3.089 & 2.329 & 45.380 &   0 & 0.70 & 60.0 & 33.5 &   0.368  & 17.265 &  0  & 0.26 & 0 & 0  \\
    & 300 & 206 &  3.678 & 2.681 & 23.578 &  13.2 & 0.93 & 52.0 & 41.1&  0.402  & 12.362 & 23.6  & 0.30 & 0    & 0     \\
    &         &         &         &         &     &      &      &      &     &      &      &      &         &    &         \\
 60 &  0 & 0 &  3.622 & 2.930 & 33.510 &   0 & 0.57 & 71.0 & 19.9 &  0.385  & 14.617 &  0  & 0.29 & 0    & 0     \\
    & 300& 189 &  4.304 & 3.928 & 31.452 &  55.3 & 0.88 & 47.4 & 26.8  & 0.371  & 14.669 & 73.3  & 0.29 & 0    & 0     \\
    & 300I& 182& 4.297 &  3.901        & 31.283 & 52.5 & 0.88 & 47.0 & 26.7 & 0.373 &    14.526 & 59.3 & 0.29 & 0 & 0 \\
    & 500 & 323& 4.447 & 3.787        & 23.944 & 27.0& 0.93 & 51.1 & 37.1 & 0.407 & 11.879 & 40.0 & 0.32 & 0 & 0 \\
    & 500I&308 & 4.440&  3.785        & 24.139 & 28.2& 0.94 & 50.8 & 37.8 & 0.405 &   11.933 & 38.3 & 0.32 & 0 & 0 \\
    &         &         &         &         &     &      &      &      &     &      &      &      &         &    &         \\
 40 &  0&  0 &  4.560 & 3.848 & 35.398 &   0 & 0.27 & 0.31 & 0.11 &  0.483  & 14.090 &  0  & 0.98 & 44.2 & 43.9  \\
    & 300 & 214 &  5.535 & 4.723 & 32.852 & 70.0 & 0.45 & 5.30 & 1.56 & 0.424  & 12.737 & 96.5 & 0.28 & 0    & 0     \\
    & 500 &     & \multicolumn{7}{|c|}{break--up limit reached} & & & & & &  \\
    & 500I& 327 & 5.700 &  4.449       & 32.085 & 165.5 & 0.49 & 5.97 & 1.76 &  & & & & \\
    &         &         &         &         &     &      &      &      &     &      &      &      &         &    &         \\

 25 & 0 &  0 &  6.594 & 5.132 & 24.209 &   0 & 0.27 & 0.31 & 0.11 &  0.688  & 16.611 &  0  & 0.46 & 10.8 & 1.73  \\
    & 300 & 239 &  8.073 & 5.566 & 21.547 & 47.0 & 0.34 & 1.75 & 0.51 & 0.633  & 11.333 & 102 & 0.98 & 25.1 & 26.4  \\
    &         &         &         &         &     &      &      &      &     &      &      &      &         &    &         \\

 20 & 0 &  0 &  8.286 & 5.168 & 19.682 &   0 & 0.27 & 0.31 & 0.11 &  0.871  & 15.745 &  0  & 0.39 & 3.61 & 0.85  \\
    & 300 & 243 & 10.179 & 5.059 & 18.074 & 82.7 & 0.31 & 1.24 & 0.35 & 0.816  & 11.787 & 0.7 & 0.41 & 3.33 & 0.85  \\
    & 300I& 240 & 10.178$^2$& 5.082 & 18.149 & 75.0 & 0.31 & 1.24 & 0.35 &         &     &   &  &  & \\
    &         &         &         &         &     &      &      &      &     &      &      &      &         &    &         \\

 15 & 0 &  0 & 11.778 & 0.000$^3$ &14.818 &   0 & 0.27 & 0.31 & 0.11 &  1.351  & 13.274 &  0  & 0.32 & 1.72 & 0.44  \\
    & 300 & 241 & 14.527 & 0.000 & 14.140 & 145 & 0.29 & 0.90 & 0.25 &1.107 & 10.203 & 0.2  & 0.36 & 2.40 & 0.58  \\
    &         &         &         &         &     &      &      &      &     &      &      &      &         &   &          \\

 12 &  0 & 0 & 16.345 & 0.000 & 11.872 &   0 & 0.27 & 0.31 & 0.11 &  2.025  & 11.599 &  0  & 0.32 & 1.80 & 0.45  \\
    & 300 & 235 & 20.377 & 0.000 & 11.500 & 183 & 0.28 & 0.77 & 0.22& 1.503  & 10.535 & 0.8 & 0.37 & 2.71 & 0.62  \\
    &         &         &         &         &     &      &      &      &     &      &      &      &         &   &          \\

  9 & 0 &  0 & 26.964 & 0.000 &  8.919 &   0 & 0.27 & 0.31 & 0.11 & 3.491  &  8.664 &  0  & 0.30 & 1.55 & 0.38  \\
    & 300 & 234 & 33.739 & 0.000 &  8.790 & 222 & 0.28 & 0.73 & 0.21 & 3.367 &  8.397 & 1.5 & 0.35 & 2.63 & 0.58  \\
    & 300I& 234 & 30.962 & 0.000 & 8.840  & 197 & 0.28 & 0.69 & 0.20 &       &   &      &    &    &  \\
    &         &         &         &         &     &      &      &      &     &      &      &      &         &    &         \\
\hline
\multicolumn{16}{l}{}\\
\multicolumn{16}{l}{$^1$ The symbol I after the velocity indicates that 
the model was computed assuming isotropic stellar winds.} \\
\multicolumn{16}{l}{}\\
\multicolumn{16}{l}{$^2$ This model was computed until the mass fraction of hydrogen at
the center is 0.000348.} \\
\multicolumn{16}{l}{}\\
\multicolumn{16}{l}{$^3$ The minimum initial mass star of O--type star is 15.9 M$_\odot$ for non--rotating models
and 17 M$_\odot$ for the rotating stellar models.}   \\

\end{tabular}
\end{center}

\end{table*}
 
The present non--rotating
12 M$_\odot$, with overshooting, presents a blue loop, while the non--rotating model of paper V,
without overshooting, shows no blue loop. 
This is at first sight surprising because, as said above, more massive He--cores tend 
to reduce the extension of the blue loops, if not suppressing them
 (Lauterborn et al. \cite{la71}; see also the discussion in Maeder \& Meynet \cite{ma89}).
As explained in paper V, this particular situation, at the limit where the loops
generally disappear, results from the higher luminosity of the 
model having the more massive He--cores (either as a result of overshooting
or as a result of rotation). The higher luminosity implies that the outer 
envelope is more extended, and is thus characterized by lower 
temperatures and higher opacities at a given mass coordinate. As a consequence, in the model
with higher luminosity, during the first dredge--up, the outer
 convective zone proceeds much more deeply in mass,
preventing the He--core to grow too much during the further evolutionary
 phases and enabling the apparition of a blue loop.

\begin{figure}[ht]
  \resizebox{\hsize}{!}{\includegraphics{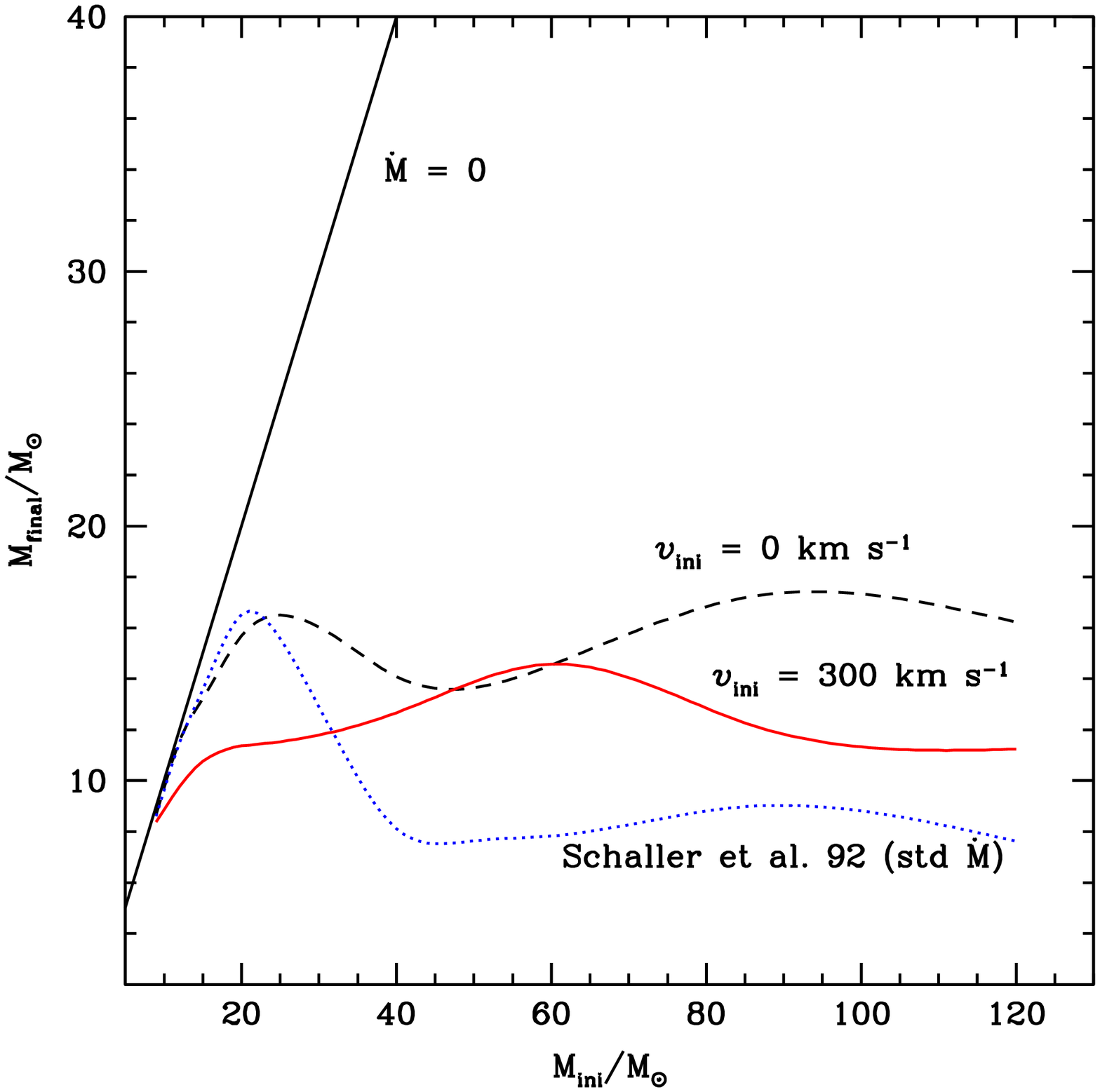}}
  \caption{Relations between the final mass versus the initial mass for solar metallicity models. 
The cases with and without rotation are indicated. 
The line with slope one, labeled by $\dot {\rm M}=0$, would correspond to a case without mass loss.
The relation obtained from
the models of Schaller et al. (\cite{Sch92}) is also shown.
}
  \label{final}
\end{figure}

The effects of rotation on the tracks have already been discussed in previous works (Heger \cite{he98}; Meynet \& Maeder \cite{MMV}), thus we shall be very brief here.
Let us just list the main effects:
\begin{itemize}
\item On and near the ZAMS, the atmospheric effects of
rotation produce a small shift of the tracks towards lower $T_{\rm eff}$.
The effective temperature considered here is an appropriate averaged $T_{\rm eff}$
 over the stellar surface (see Meynet \& Maeder \cite{MMI}).

\item Tracks with rotation become more luminous than for no rotation.

\item For stars with initial masses inferior to $\sim$ 30 M$_\odot$, the MS bandwidth is enlarged by rotation. This is a consequence
of the larger He--core produced during the H--burning phase in rotating models. As a numerical example,
in the 20 M$_\odot$ model with $v_{\rm ini}$ = 300 km s$^{-1}$, the He--core at the end of the MS is about 38\% more massive than in its 
non--rotating counterpart. In the models of paper V the enhancement
due to rotation was only 20\%. Likely most of the difference
results from the inclusion here of a small overshooting. 
In the present models, the core enhancement
due to rotation adds its effects to an already more extended core, having a higher central
temperature, and thus this larger core reacts more strongly to further supplies of fuel
by rotational diffusion. 

\item For initial masses superior to about 50 M$_\odot$, the rotating tracks 
show drastically reduced MS bandwidths.
This was already the case in Meynet \& Maeder (\cite{MMV}).
For instance, the reddest point reached at the end of the MS phase by the 
60 M$_\odot$ non--rotating stellar model is log $T_{\rm eff} \sim$ 4.15 , while
for the $v_{\rm ini}$ = 300 km s$^{-1}$ model, it is $\sim$ 4.6 , due to the fact 
that the surface He--enhancement leads to a bluer track, as a consequence of the reduced opacity. 
In this mass range, the rotating stellar models enter into the WR phase during 
the H--burning phase (see Sect. 5). The MS phase, identified here with the 
OV--type star phase, is thus considerably reduced.
Since the stars, here as in any  mass range, have different  initial rotational velocities, 
one expects to find in the HR diagram stars corresponding to a
variety of tracks with various initial velocities. Particularly, in the upper hot
part of the HR diagram, for log $L/L_\odot$ superior to about 5.8--5.9 
and log $T_{\rm eff}$ between 4.7 and 4.2, one may find in the same area of the HR
diagram stars in very different evolutionary stages, which may also correspond to
different types of massive stars:  MS O--type stars, Of stars, transition stars,
blue supergiants and WR stars.  

\end{itemize}

For higher rotation velocities, the actual masses at the end of both the H-- and He--burning phases
become smaller (cf. Tables~\ref{tbl-1}). For an initial velocity of 300 km s$^{-1}$ and for stars
with masses below $\sim$ 60 M$_\odot$, 
the actual masses at the end of the core H--burning phase are decreased by less than 11\% with respect
to the non--rotating models. For higher initial mass models, the decreases are much more important, because
these stars enter the WR phase already at an early stage during the H--burning phase (see Sect. 5).

The evolutionary masses of a star can be determined by searching the mass 
of the evolutionary track passing through
its observed position in the HR diagram.
As can be seen from Fig.~\ref{hr}, the mass determined from the non--rotating evolutionary track is always
greater than the one deduced from rotating models.  Thus the use of non 
rotating tracks tends to overestimate the mass, and
this might be a cause of a large part of the long--standing mass discrepancy problem
 (see e.g. Herrero et al. \cite{He20}).

Fig.~\ref{final} shows the final masses obtained for different initial masses and velocities. 
One notes that the new mass loss rates used in the present computations
are smaller than those used
in the stellar grids of Schaller et al. (\cite{Sch92}).
For stars with initial masses 
above 40 M$_\odot$,
rotating models produce 
final masses between 11.3 and 14.7. For comparison, non--rotating models   
end with final masses between 14.1--17.3 M$_\odot$. Interestingly, the average mass
of WC stars, estimated on the basis of the 6 WC stars
belonging to double--line spectroscopic binaries, is 12 $\pm$ 3 
M$_\odot$ (van der Hucht \cite{Hu01}), nearer to the rotating model results than the non--rotating
ones. However we cannot discard the possibility that for these stars mass
transfer in a Roche lobe overflow may have occurred, in which case this
observation cannot be used to constrain single star models.
The fact that the models with and without rotation
near 50 M$_\odot$ lead to about the same final masses, despite the
increased mass loss rates for rotating models, is due to the differences in the
evolutionary stages followed by each model (see Sect.~ 5).

\begin{figure}[ht]
  \resizebox{\hsize}{!}{\includegraphics{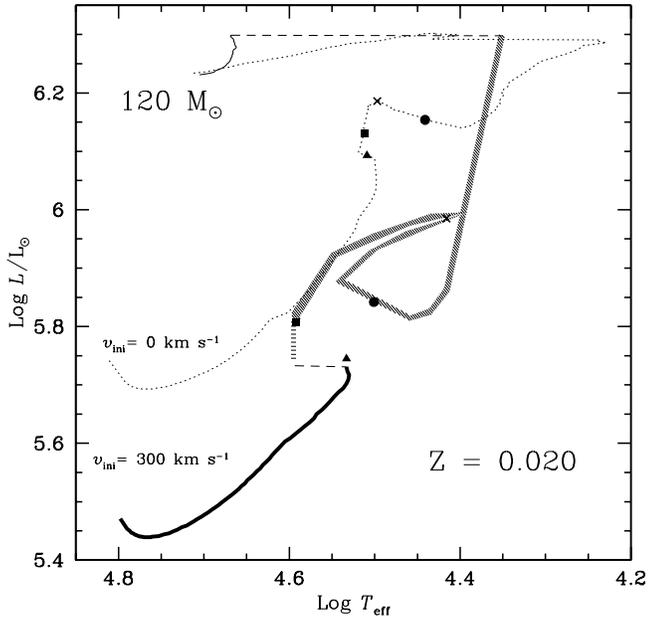}}
  \caption{Evolutionary tracks for a non--rotating (light dotted--line) and a rotating 120 M$_\odot$ stellar model
at solar metallicity. The circles along the tracks indicate the end of the eWNL phase, the squares
the end of the eWNE phase, and the triangles, the beginning of the WC phase. The WN/WC stars lie along
the portion of the tracks between the squares and the triangles. The crosses indicate the end of the central H--burning phase.
}
  \label{hr120}
\end{figure}

Table~\ref{tbl-1} presents some properties of the models. Columns 1 and 2 give the initial mass and the initial velocity $v_{\rm ini}$ respectively.
The mean equatorial velocity $\overline{v}$ during the MS phase is indicated in column 3.
This quantity is defined as in paper V. 
The H--burning lifetimes $t_H$, the lifetimes $t_{\rm O}$ as an O--type star on the MS 
(we assumed that O--type stars have an effective temperature superior to about
31 500 K as suggested by the new effective temperature scale given by Martins et al. \cite{Mart02}),
the masses M, the equatorial velocities $v$, the helium surface abundance $Y_s$ and the 
surface ratios (in mass) N/C and N/O at the end of the H--burning phase are given in columns 4 to 10.
The columns 11 to 16 present some characteristics of the stellar models at the end of the He--burning phase,
$t_{He}$ is the He--burning lifetime.
The letter I attached to the value of the initial velocity (column 2) indicates
that the model was calculated assuming isotropic stellar winds. These models were
calculated until the end of the MS phase, except in the case of the 60 M$_\odot$
model for which the evolution was pursued until the end of the core He--burning
phase.

From Table~\ref{tbl-1} one sees that for $Z=0.020$ the lifetimes are 
increased by about 15--25\% when the initial rotational velocity 
on the MS increases from 0 to 300 km s$^{-1}$. Similar increases
were found in paper V. The lifetimes of O--type stars are increased by similar amounts. 
As we shall see below, the increase due to rotation of the WR lifetimes is much greater
and thus rotation predicts larger number ratios of WR to O--type stars 
than non--rotating stellar models. 

\begin{figure} [ht]
\resizebox{\hsize}{!}{\includegraphics{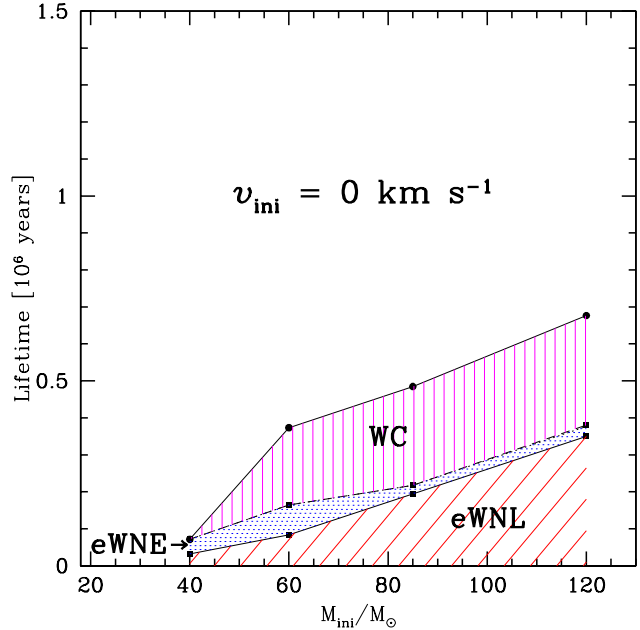}}
\caption{Lifetimes of WR stars of different initial mass for
non--rotating stellar models.
The durations of the WR sub-phases are also indicated. There is no
transition WN/WC phase.}
  \label{tma}
\end{figure}

Except for the models with initial masses superior to 85 M$_\odot$,
the He--burning lifetimes are decreased by rotation. Thus the ratio of the He--burning lifetimes
to the H--burning lifetimes are in general decreased by rotation. 
Typically for the 20 M$_\odot$ model, this ratio
passes from 0.11 to 0.08 when the initial velocity considered increases
 from 0 to 300 km s$^{-1}$. The particular behaviour of the extremely massive 
 stars results from the lower luminosities in the WC stage, which
favour longer He--lifetimes.

The anisotropy of the wind induced by rotation has little impact when the initial
velocity is equal or inferior to 300 km s$^{-1}$. This can be seen by comparing
the data presented in Table~\ref{tbl-1} for the two 60 M$_\odot$ models with
$\upsilon_{\rm ini}$ = 300 km s$^{-1}$ computed with and without the account of
the anisotropy. They are indeed very similar. On the other hand, as has been shown by
Maeder (\cite{Ma02}), for higher initial velocities, in the mass range
between 20 and 60 M$_\odot$, the anisotropies of the mass loss 
may significantly influence the evolution of the stellar velocities 
and may lead many stars to break--up.

\section{Rotation and the properties of Wolf--Rayet stars}

\begin{figure}[ht]
  \resizebox{\hsize}{!}{\includegraphics{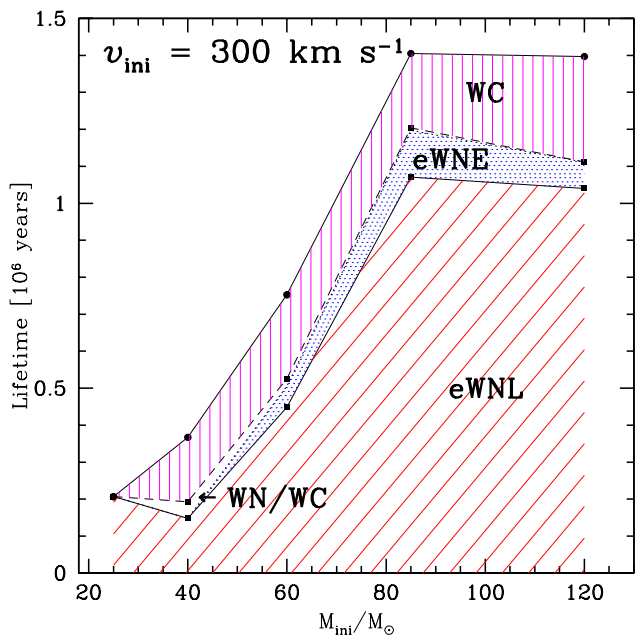}}
  \caption{Lifetimes for rotating stellar models in the WR subphases.
}
  \label{tma2}
\end{figure}

Here we consider the following question:
what are the effects of rotation on the evolution of massive single 
stars into the Wolf--Rayet phase ? This subject has been discussed by Maeder (\cite{Maeder87}),
Fliegner and Langer (\cite{Fl95}), Maeder \& Meynet (\cite{araa}) and Meynet (\cite{Me00}).
We present here for the first time an extensive grid of models, accounting for the effects
of the wind anisotropy, going through the WR phase. It is important to cover the whole
mass range of star progenitors of WR stars in order to quantitatively assess the importance of rotation
in the interpretation of the WR number statistics (see Sect. 6 below).

Let us reconsider the criteria which have been chosen 
to decide when a stellar model enters into the WR phase. 
Ideally, of course, the physics of the models should decide 
when the star is a WR star. However our poor knowledge of the atmospheric physics 
involved, as well as the complexity of models
coupling stellar interiors to the winds, are such that this 
approach is not yet possible.
Instead, it is necessary to adopt some empirical criteria
for deciding when a star enters the WR phase. 
In this work the criteria are the following: 
the star is considered to be a WR star when
$\log T_{\rm eff} > 4.0$
and the mass fraction of hydrogen at the surface is $X_{\mathrm{s}} < 0.4$. 
Reasonable changes to these values (for instance adopting $X_{\mathrm{s}} < 0.3$ 
instead of 0.4) do not affect the results significantly.

The correspondence between the WR--subclasses and the surface abundances is established as follows:
the stars enter the WR phase as WN stars {\it i.e.} as stars presenting strong He and N enrichments at their surface
and strong depletion in carbon and oxygen as expected from CNO processed material.
We call eWNL stars all the WN stars showing hydrogen at their surface 
(``e'' stands for evolutionary, as recommended by Foellmi et al. \cite{Fo03}).
The  eWNE stars, on the other hand, are WN stars showing no hydrogen at their surface.
The transitional WN/WC phase is characterized by the simultaneous presence at the surface of both
H-- and He--burning products (typically N and C or Ne). Here we suppose 
that this transition phase begins when the
mass fraction of carbon at the surface becomes superior to 10\% 
of the mass fraction of nitrogen. The phase is supposed
to end when the mass fraction of nitrogen at the surface becomes inferior 
to 10\% of the mass fraction of carbon.
The WC phase corresponds to the stage where He--burning products appear at the
 surface. This phase begins when the transition WN/WC phase ends.

\begin{figure}[ht]
  \resizebox{\hsize}{!}{\includegraphics{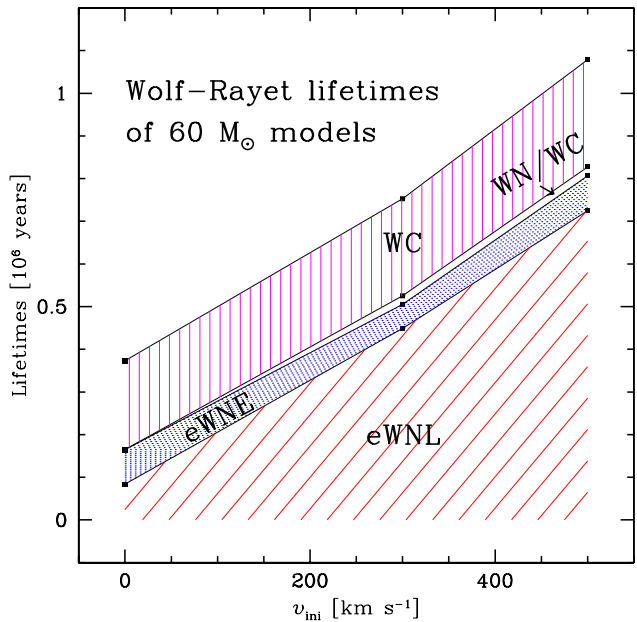}}
  \caption{Durations of the WR phases and of the WR sub-phases for stars of 60 M$_\odot$
with various initial velocities.
}
  \label{twrvi}
\end{figure}

Figure~\ref{hrkip} shows the evolutionary tracks and the evolution
 of the structure in a non--rotating and a rotating 60 M$_\odot$ model. 
The rotating model has a time--averaged equatorial velocity on the Main 
Sequence (MS) of about 190 km s$^{-1}$, 
not far from the mean equatorial velocity observed for O--type stars. Inspection of 
Fig.~\ref{hrkip} shows the following features:
\vskip 2mm
\begin{itemize}
\item The non--rotating 60 M$_\odot$ star goes through an intermediate Luminous Blue Variable (LBV) phase
before becoming a WR star, while the rotating model skips the 
LBV phase and goes directly from the O--type star 
phase into the WR phase. One also notes  that during the O--type star phase, 
the rotating track is bluer than its non--rotating counterpart. 
This is due to the diffusion of helium into the outer radiative zone (see Heger \& Langer \cite{He00};
Meynet \& Maeder \cite{MMV}).  This fact may slightly change the number of O--type stars deduced 
from the number of ionizing photons, a technique used to estimate the number 
of O--type stars from the spectral analysis of the integrated light of
distant starburst regions (e.g. Schaerer \cite{Sc96}).

\item There is no luminosity difference during the WC phase
 between the rotating and the non--rotating models. 
Since these stars follow a mass--luminosity relation (Maeder \cite{Maeder87}; Schaerer \& Maeder \cite{Sc92}),
this implies that the masses of these two stars at this stage are about equal (see Fig.~\ref{final}). 
At first sight this may be surprising. One would expect the rotating star to lose
 more mass than its non--rotating counterpart. 
However, in this particular case there are some compensating effects. Indeed,  
the non--rotating model enters into the WR phase later, but becomes redder during the MS phase and then becomes a LBV star.
During these two last phases, the star undergoes strong mass loss (see the left panels of Fig.~\ref{hrkip}). 
The rotating model, on the other hand, enters 
the WR phase at an earlier stage, reaching at the end an identical mass to the non--rotating model.
In general, however, the final masses obtained from the rotating models are smaller than those obtained from the non--rotating ones
(see Fig.~\ref{final}), 
leading to lower luminosities during the WC phase (see Fig.~\ref{hr120}).

\item From the lower panels of Fig.~\ref{hrkip}, we see that the non--rotating star enters 
the WR phase at the beginning of the core He--burning phase
with a mass of about 27 M$_\odot$. 
Nearly the whole H--rich envelope has been removed by stellar winds during the previous phases 
(more precisely at the end of the MS phase and during the LBV phase). 
In this case, the main mechanism for making a WR star is 
mass loss by stellar winds. For the rotating model, on the other hand,
the entry into the WR phase occurs at an earlier stage 
(although not at a younger age!), while the star is still burning hydrogen in its core. 
The mass of the star at this stage is about 45 M$_\odot$
and a large H--rich envelope is still surrounding the convective core.
The main effect making a WR star in this second case is
rotationally induced mixing (Maeder \cite{Maeder87}; Fliegner \& Langer \cite{Fl95}; Meynet \cite{Me00}).

\end{itemize}
\vskip 2mm

In Fig.~\ref{hr120}, the evolutionary tracks for a non--rotating and a rotating 120 M$_\odot$ stellar model
are shown. The beginning and the end of the various WR subphases are indicated
 along the tracks. 
The range of luminosities spanned by the rotating eWNL stars is 
considerably increased. This directly results from the point underlined just above, {\it i.e.}
a rotating star enters the WR phase still with a large H--rich envelope. This considerably
increases the duration of the eWNL phase and the star will cover during this stage a greater
range of masses and luminosities. The rotating models also enter the further
WR subclasses with smaller actual masses and lower luminosities.
From Fig.~\ref{hr120}, one sees that 
the rotating eWNE, WN/WC and WC stars have luminosities decreased by about a factor of two.

\begin{figure}[ht]
  \resizebox{\hsize}{!}{\includegraphics{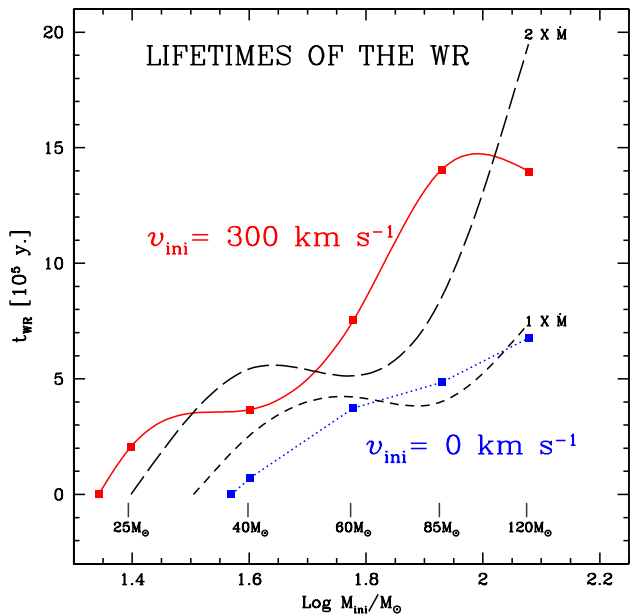}}
  \caption{Lifetimes of Wolf--Rayet stars from various initial masses at solar metallicity.
The continuous line shows the results from the present rotating models. The WR lifetimes of
the non--rotating
models from the present work are shown by the dotted line. The short-- and long--dashed curves
show the results obtained from the
models of Schaller et al. (\cite{Sch92}) and 
Meynet et al. (\cite{Mey94}) with normal (1 X $\dot M$) and enhanced mass loss rates (2 X $\dot M$)
respectively.}
  \label{twr}
\end{figure}

The differences between the behaviours of the rotating and 
non--rotating models have important consequences for the duration of the WR phase as a whole
and for the lifetimes spent in the different WR subtypes, as can be appreciated from Figs.~\ref{tma} and
\ref{tma2}.
\vskip 2mm
\begin{itemize}
\item Firstly, one notes that the WR lifetimes are enhanced. Typically 
for the 60 M$_\odot$ model, the enhancement by rotation amounts to nearly a factor of 2.

\item  Second, the eWNL phase is considerably lengthened as explained just above.
The duration of this phase increases with the initial velocity as can be seen from Fig.~\ref{twrvi}.
When the initial velocity passes from 300 to 500 km s$^{-1}$ 
(or when the time--averaged equatorial velocity during the O--type phase goes 
from 190 to 330 km s$^{-1}$), the eWNL phase duration is increased by a factor of about 1.7.
The durations of the eWNE and WC phases are, on the other hand, much less affected.

\item  Thirdly, in the rotating stellar model a new phase of modest,
but of non--negligible duration, appears: the so--called transitional WN/WC phase. 
This phase is characterized by the simultaneous presence at the surface of both
H-- and He--burning products. The reason for this
is the shallower chemical gradients that build up inside the rotating models.
These shallower gradients inside the stars also produce a smoother evolution of the surface abundances 
as a function of time (or as a function of the remaining mass, see Figs.~\ref{xi60} and \ref{xi40}). 
For a transitional WN/WC phase to occur, it is necessary to have---for a sufficiently long period---both 
a He--burning core and a CNO--enriched envelope.
In the highest mass stars, mass loss removes too rapidly the CNO--enriched envelope 
to allow a transitional WN/WC phase to appear. 
In the low mass range, the time spent in the WR phase is 
too short and the H--rich envelope too
extended to allow He--burning products to diffuse up to the surface. 
Consequently, the transitional WN/WC phase only appears in the mass range between
$\sim$ 30 and 60 M$_\odot$ for $v_{\rm ini}=300$ km s$^{-1}$.

\item Finally, the minimum mass for a star to become a WR star (through the single star channel) 
is lowered by rotation. In the present case, the minimum mass is reduced from about 
37 M$_\odot$ for $v_{\rm ini}=0$ km s$^{-1}$
to about 22 M$_\odot$ for $v_{\rm ini}=300$ km s$^{-1}$.

\end{itemize}
Of course, this applies to solar metallicity. At very different $Z$, the effects of
rotation on the various subphases may be different.
It is interesting to compare the present WR--lifetimes with those we obtained with our
previous non--rotating stellar models (Meynet et al. \cite{Mey94}; Maeder and Meynet \cite{MaeMey94}).
This is done in Fig.~\ref{twr}. We see that the present non--rotating models follow more or less
the behaviour we obtained with our normal mass loss rate models of 1994, although, due to smaller mass loss rates
in the present grid, the WR lifetimes are in general reduced and the minimum mass for WR star formation from single stars is enhanced.
Our rotating models, on the other hand, reproduce qualitatively the changes brought by an enhancement of the mass loss rates: an increase of the WR lifetimes and a decrease of  
the minimum mass for a single star to go through a WR phase.

From the above considerations, one can conclude that, compared
to non--rotating models, stellar models including rotation will
predict higher values of: (a) the numbers of WR stars relative to O--type stars, 
(b) the relative number of WN to WC stars, and (c) the relative number of transition WN/WC stars to WR stars. As shown in the next Section, the predictions of the rotating  models
 are in better agreement with the observed number ratios at solar metallicity.

\section{Comparison with the observed properties of Wolf--Rayet star populations}

Following Maeder (\cite{Maeder91}) and
Maeder \& Meynet (\cite{MaeMey94}), one can easily estimate the theoretical
number ratio of WR to O--type stars in a region of constant star formation.
This ratio is simply given by the ratios of the mean lifetime of a WR star to that
of an OV--type star. The averages are computed using an  initial mass
function as the weighting factor over the respective mass intervals considered.
Assuming a Salpeter Initial Mass Function slope of 1.35, considering the O--type and WR star lifetimes 
given in tables~\ref{tbl-1} and \ref{tblwr}, we obtain the predicted ratios given at the bottom
of this last table.

\begin{table}[ht]
\caption{Lifetimes of Wolf--Rayet star from various initial masses and velocities
and predictions for different number ratios (see text). The velocities are in km s$^{-1}$ and the lifetimes in Myr.} 
\label{tblwr}
\begin{center}\scriptsize
\begin{tabular}{ccccccc}
M   & $v_{\rm ini}$        & $t_{WR}$ & $t_{eWNL}$ & $t_{eWNE}$ & $t_{WN/WC}$ & $t_{WC}$  \\
    &                   &          &           &           &             &           \\
\hline
    &                   &          &           &           &             &           \\
120 & 0                 & 0.6767   & 0.3499    & 0.0265    & 0.0038      & 0.2965    \\ 
    & 300               & 1.3969   & 1.0402    & 0.0706    & 0.0010      & 0.2851    \\
    &                   &          &           &           &             &           \\
 85 & 0                 & 0.4846   & 0.1941    & 0.0231    & 0.0000      & 0.2674    \\ 
    & 300               & 1.4050   & 1.0707    & 0.1263    & 0.0065      & 0.2015    \\
    &                   &          &           &           &             &           \\
 60 & 0                 & 0.3732   & 0.0838    & 0.0808    & 0.0000      & 0.2086    \\
    & 300               & 0.7527   & 0.4489    & 0.0562    & 0.0202      & 0.2274    \\
    & 300I              & 0.7779   & 0.4734    & 0.0346    & 0.0343      &    0.2356 \\
    & 500               & 1.0787   & 0.7250    & 0.0816    & 0.0206      &    0.2515 \\
    & 500I              & 1.0703   & 0.7312    & 0.0628    & 0.0211      &   0.2552  \\
    &                   &          &           &           &             &           \\
 40 & 0                 & 0.0714   & 0.0317    & 0.0397    & 0.0000      & 0.0000    \\
    & 300               & 0.3670   & 0.1478    & 0.0000    & 0.0447      & 0.1745    \\
    &                   &          &           &           &             &           \\
 25$^1$ & 0                 & 0.0000   & 0.0000    & 0.0000    & 0.0000      & 0.0000    \\
    & 300               & 0.2068   & 0.2068    & 0.0000    & 0.0000      & 0.0000    \\
    &                   &          &           &           &             &           \\
\hline
    &                   &          &           &           &             &           \\
 \multispan{7}\hfill Predicted number ratios \hfill \\
    &                   &          &           &           &             &           \\
 $v_{\rm ini}$        & ${\rm WR} \over {\rm O}$       & ${\rm eWNL}\over {\rm WR}$   
 & ${\rm eWNE}\over {\rm WR}$   & ${\rm WN/WC}\over {\rm WR}$  & ${\rm WC}\over {\rm WR}$  & ${{\rm WC}\over {\rm WN}}^{2}$     \\
                      &          &           &           &             &        &           \\
\hline
                      &          &           &           &             &        &           \\
    0                 & 0.02     & 0.35      & 0.16      & 0.00        & 0.49   &  0.97     \\ 
                      &          &           &           &             &        &           \\
    300               & 0.07     & 0.66      & 0.05      & 0.04        & 0.25   &  0.35     \\
                      &          &           &           &             &        &           \\
\hline   
                     &          &           &           &             &        &           \\                       
\multispan{7}$^1$ M$_{min}$ for WR star formation from a single star is 37 and 22\hfill \\
\multispan{7}for $\upsilon_{\rm ini}=0$, 300 km s$^{-1}$ respectively.\hfill \\
\multicolumn{7}{l}{}\\
\multispan{7}$^2$ WC/WN = WC/(WNL+WNE). \hfill \\ 
\end{tabular}
\end{center}

\end{table}

\begin{figure}[tbp]
  \resizebox{\hsize}{!}{\includegraphics[angle=-90]{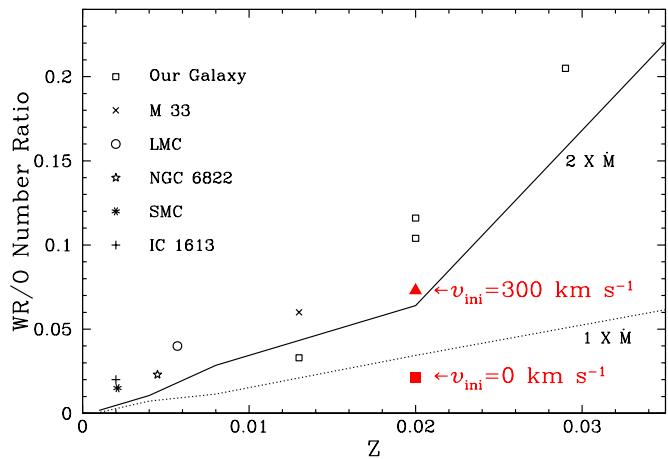}}
  \caption{Variation of the number ratios of Wolf--Rayet stars to O--type stars as a function of the metallicity.
The observed points are taken as in Maeder \& Meynet (\cite{MaeMey94}).
The dotted and continuous lines show the predictions of the models
of Schaller et al. (\cite{Sch92}) and Meynet et al. (\cite{Mey94}) with normal and enhanced mass loss rates respectively.
The black square and triangle show the predictions of the present non--rotating and rotating stellar models
respectively. 
}
  \label{wro}
\end{figure}

\begin{figure}[tpb]
  \resizebox{\hsize}{!}{\includegraphics[angle=-90]{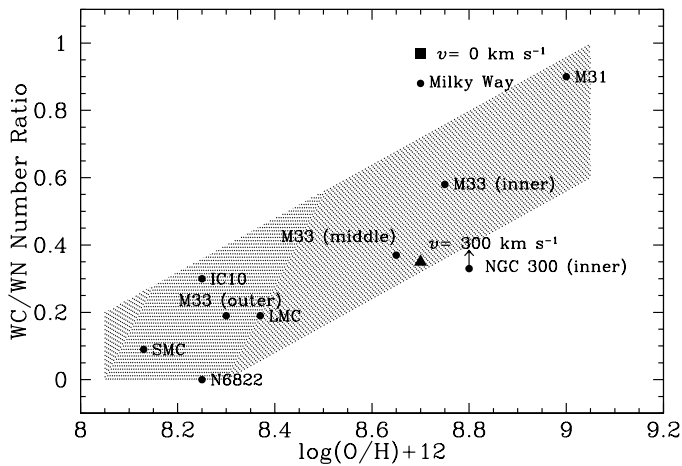}}
  \caption{Variation of the number ratios of WN to WC stars as a function of metallicity.
The black circles are
observed points taken  from Massey \& Johnson (\cite{Mas98} and see references therein), except
for the SMC (Massey \& Duffy \cite{Mas01}), for NGC 300 (Schild et al. \cite{Sch02})
and for IC10, for which we show the estimate from Massey \& Holmes (\cite{Mas02}) .
The black square and triangle show the predictions of the present non--rotating and rotating stellar models
respectively.}
  \label{massey}
\end{figure}

Comparisons with observed number ratios are shown in Figs.~\ref{wro} and \ref{massey}.
Obviously, 
unless most of the WR stars at solar metallicity are formed through Roche 
lobe overflow in a close binary system, which seems quite unrealistic 
(see the discussion in Maeder \cite{Maeder96}; see also Foellmi et al. \cite{Fo03}),
 models with rotation are in much better
agreement with observations. Of course a more refined estimate of the effects 
of rotation will require account of the initial distribution of the 
rotational velocities. 
However, since the initial velocity chosen here corresponds to a mean 
rotational velocity on the Main Sequence almost equal to the observed mean,
we can hope that the present result is not too far from what one would 
obtain with a more complete analysis. The conclusion is that
at solar metallicity, rotating models are in much better agreement
with the observed statistics of WR stars than the non--rotating models.
Interestingly a similar conclusion could be reached based on the results
of our paper V, and this shows that the present results depend weakly
on overshooting or on the inclusion of the effects of horizontal turbulence
in the coefficient of  shear diffusion.

In Fig.~\ref{massey}, the predicted values for the WC/WN number ratios are compared
with observations. One sees that the WC/WN ratio increases with the metallicity
along a relatively well defined relation.
The observed point
for the solar neighborhood is however well above the general trend. According to Massey 
(\cite{Mas03}) this may result from an underestimate of the number of WN stars.

Comparing with the theoretical predictions at solar metallicity, one obtains that
the non--rotating models predict too few WN stars, while
the rotating models give a value in the observed trend. 
This confirms the conclusion already reached above on the basis
of the WR/O number ratio: rotating models
much better account for the characteristics of the WR populations than
non--rotating models.

There are 10  WN/WC stars listed in the VIIth catalogue of galactic WR stars of van der Hucht (\cite{Hu01}). 
Compared to the total number of known galactic WR stars (226) this represents
a fraction of 4.4\%. The present models predict for this fraction
a value equal to 4\% in good agreement with the observed ratio.
We recall that models without mild mixing processes in radiative
zones are unable to predict correctly the number of transition WN/WC stars.
This was first shown by Langer (\cite{La91}).

In future work, it has to be checked
by further computations at different metallicities that 
the rotating models, with reduced mass loss rates due to clumping,
will account for the observed variation of the number ratios WR/O and WC/WN 
with metallicity.

\section{Surface abundances in WR stars}

\begin{figure}[ht]
  \resizebox{\hsize}{!}{\includegraphics{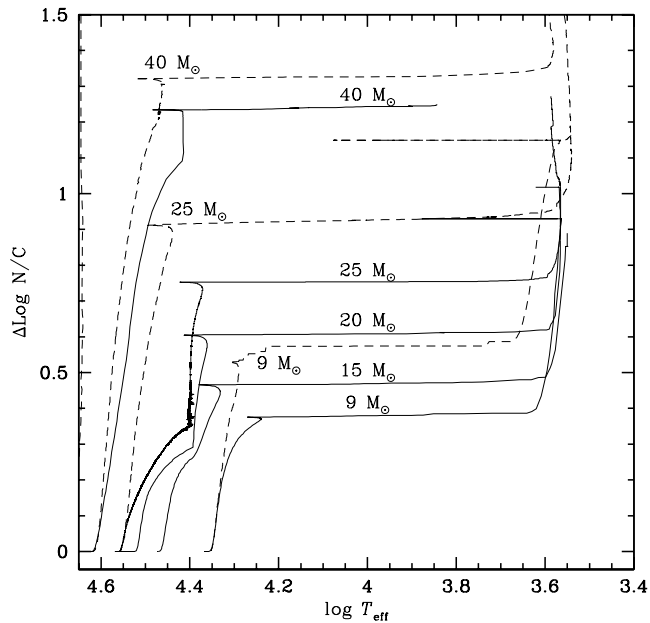}}
  \caption{Evolution of the N/C ratios (in number) as a function of the
  effective temperature at the surface of various rotating stellar models. The
  N/C ratios are normalized to their initial values. The
  continuous lines show the results for the present rotating models ($\upsilon_{\rm ini}$ = 300 km s$^{-1}$),
  the dashed--lines show the evolutionary tracks for some models of paper V. For purpose of clarity, only
  a part of the 40 M$_\odot$ tracks are shown.}
  \label{nc}
\end{figure}

We have already presented and discussed the effects of
rotation on the surface abundance of massive stars at solar composition
(Meynet \& Maeder \cite{MMV}). The main point is the enhancement 
of He and N  during the MS phase due to the internal transport by shear mixing.
However, we note that the enhancements in He and CNO products found here
are in general smaller that those found in Meynet \& Maeder (\cite{MMV}). 
The situation is illustrated in 
Fig.~\ref{nc} which compares the evolution of the N/C ratios
for models with masses between 9 and 40 M$_{\odot}$ of the  above quoted paper  and of 
the present work. We see that a difference in the 
enhancement of N/C is created during MS evolution and that the present ratios
are smaller  on average by a factor of about 1.5 for masses below 25 M$_\odot$.
This is  due to the fact that 
in the present models we are accounting for the effects 
of the  horizontal turbulence on the shear
mixing according to the study by Talon \& Zahn (\cite{TaZa97}). The horizontal
diffusion substantially reduces the effects of the shear mixing.  In the higher mass range, the difference
between the two sets of models is smaller, as can be seen by comparing the two 40
M$_\odot$ tracks in Fig.~\ref{nc}. In this domain of masses, the large effects of mass
loss by stellar winds are added to those of rotational mixing.
Close comparisons with
observations will be undertaken  on the basis of these two sets of models to  see 
whether the treatment  by Talon \& Zahn leads to 
a  better agreement or not. However, the problem is intricate both
observationally, because the scatter of the observations is large,
and also theoretically  because other effects may enter into the game, such as 
magnetic fields (Spruit \cite{spruit02}) or a higher horizontal
turbulence. Let us also
recall that the enhancements in He and N are much larger at lower 
metallicities $Z$ (Maeder \& Meynet \cite{MMVII}), because stars with lower $Z$
have in general  steeper $\Omega$--gradients, which favour internal mixing.

\begin{figure}[t]
  \resizebox{\hsize}{!}{\includegraphics{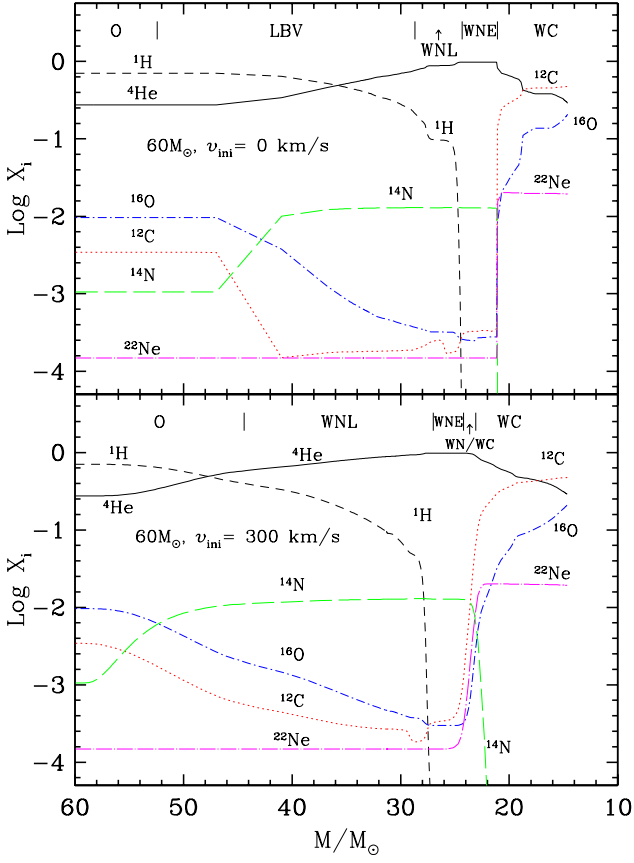}}
  \caption{Evolution as a function of the actual mass of the abundances (in mass fraction) at the surface of a non--rotating  
(upper panel) and a rotating (lower panel) 60 M$_\odot$ stellar model.
}
  \label{xi60}
\end{figure}

Fig.~\ref{xi60} shows the evolution of the surface abundances in a rotating and 
a non--rotating model with initial masses of 60 M$_{\odot}$. The progressive changes
of the abundances of CNO elements between the initial cosmic values  
and the values of the nuclear equilibrium of the
CNO cycle are much smoother  for the rotating model than for the case without
rotation. This is due to the internal mixing which makes flatter internal
chemical gradients for rotating models.  The rotational mixing also makes the change
of abundances to occur  much earlier in the peeling--off process.
 This is also true for the changes of 
H and He. Another significant difference for the  60 M$_{\odot}$ model with rotation is the absence
of an LBV phase as shown in Sect. 5. Without rotation, most of the change of CNO abundances
occurs in the LBV phase, while in the case with rotation, the transition 
from cosmic to CNO equilibrium abundances occurs partly during the MS phase
and partly during the eWNL phase which immediately follows it. We note that
the nuclear equilibrium CNO values are essentially model independent as already stressed
a long time ago (Smith \& Maeder \cite{SmithM91}). This is true whether H is still 
 present or not.

 We see that
 at the end of the  assumed eWNE phase, the transition to the WC abundances 
is very sharp in the absence of rotational mixing. This is because
 there is a strong chemical discontinuity in the classical models due to the fact
 that the convective core is usually growing during some part of the 
 He--burning phase (Maeder \cite{Maeder83}). In the rotating case, the transitions are smoother,
 so that there are some stars observed with abundances in the  transition regime,
 with simultaneously some $^{12}$C and some $^{14}$N present.
 As stated above, these   correspond to the so--called  WN/WC stars
 (Conti \& Massey \cite{Co89}; Crowther et al. \cite{Cr95}).
 These transition stars are likely to also
have some  $^{22}$Ne excess. Since the attribution of spectral types
is a complex matter, it may even be that some of the stars in the 
transition stage are given a spectral type eWNE or WC, thus we might well
have a situation where a WN star would have some  $^{22}$Ne excess
or a WC star would still have some $^{14}$N present.

\begin{figure}[ht]
  \resizebox{\hsize}{!}{\includegraphics[angle=-90]{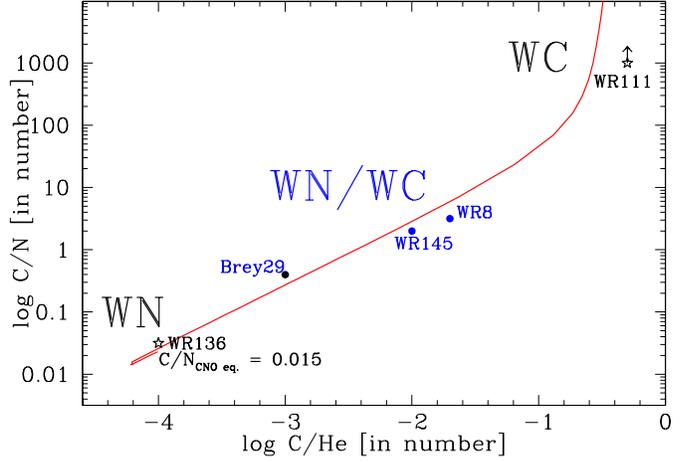}}
  \caption{Evolution of the abundance ratio C/N at the surface of
  a 40 M$_\odot$ stellar model with $\upsilon_{\rm ini}$ = 300 km s$^{-1}$
  as a function of the ratio C/He. Only the beginning of the WR phase
  is shown. The equilibrium value of the C/N ratio is around 0.015.
  Observations as given in the paper by Crowther et al. (\cite{Cr95}) are
  indicated.}
  \label{wnc}
\end{figure}

Fig.~\ref{wnc} compares the predicted relation between the C/N and C/He ratios
with some recent observations by Crowther et al. (\cite{Cr95}). 
During the eWNL phase, C/He and C/N decreases, but remain around
the CNO equilibrium value equal to 0.015.
Then during the transitional WN/WC phase, there is a huge increase
of carbon, while nitrogen and helium remain constant. This is the reason
why the lower part of the curve shows that when C/N changes by a factor 
$10^3$, C/He changes by the same amount.
Then, when the star evolves towards the WC stage, C/He changes only slowly, while 
C/N grows very rapidly, since N quickly disappears. The comparison with the observations 
by Crowther et al. (\cite{Cr95}) shows a good agreement. However this agreement may be somewhat
misleading here, in the sense that a 60 M$_\odot$ non--rotating
model would follow a similar track in the C/N versus C/He plane. 
The good correspondence here is not a real constraint on the models. What is more
constraining is the expected number of stars in the transitional WN/WC phase. As seen
in Sect. 5, the rotating models in this respect are much better than the non--rotating ones.  

In the WC stage, the abundances of chemical elements are very much model
dependent as was first shown by Smith \& Maeder (\cite{SmithM91}). 
This is because in the WC stars, we see the products of partial He-burning.
At the entry in the WC phase, the $^{12}$C and  $^{16}$O abundances are lower, 
and that of He is higher  in the models with rotation, due to their smooth chemical gradients.
Also, the fraction of the WC phase spent with lower C/He and O/He ratios is longer 
in models with rotation. However, rotation does not affect the high $^{22}$Ne abundance
during the WC phase. This is a consequence of the fact that most of this $^{22}$Ne results from the transformation of the $^{14}$N produced by the CNO cycle in the previous H--burning core. The value of the $^{14}$N and therefore that of the $^{22}$Ne is fixed by the
characteristics of the CNO at equilibrium, which in turn depends on the nuclear physics and not on the pecularities of the stellar models. It is interesting
to mention here that the high overabundance of $^{22}$Ne at the surface of the
WC star predicted by the models is confirmed by the observations (Willis
\cite{Wi99}; Dessart et al. \cite{De00}). In the core, the abundance of $^{22}$Ne 
will stay at this equilibrium value until eventually it is turned into $^{25,26}$Mg 
with production of neutrons (through $^{22}$Ne($\alpha$,n)$^{25}$Mg)
and of s--elements.

\begin{figure}[t]
  \resizebox{\hsize}{!}{\includegraphics{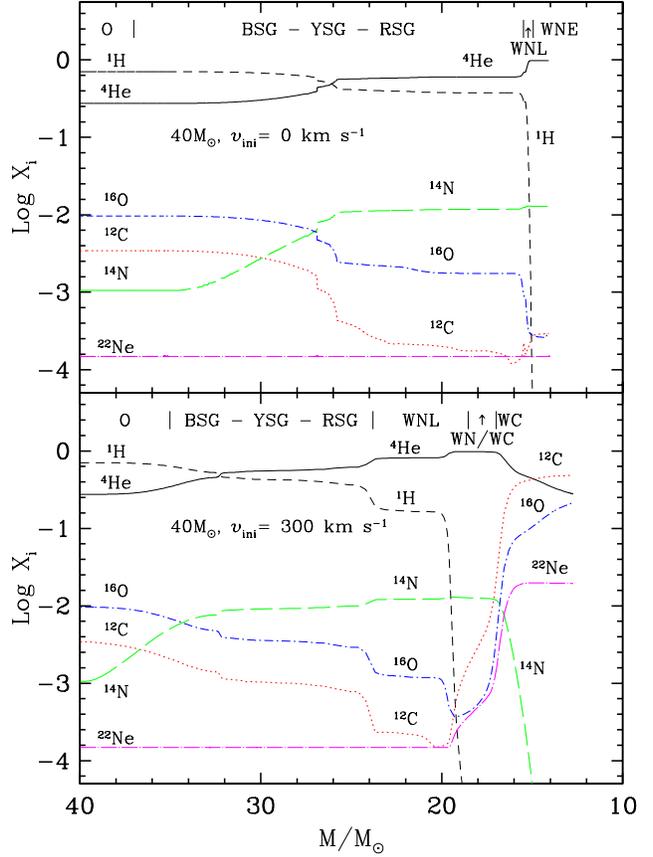}}
  \caption{Same as Fig.~\ref{xi60} for 40 M$_\odot$ stellar models.
}
  \label{xi40}
\end{figure}

Fig.~\ref{xi40} shows the same for models of 40 M$_{\odot}$ with and without rotation.
The enhancements of He and N only appear in the blue  and yellow supergiant stage
in the case without rotation, while they occur during the MS and blue supergiant phase
when rotation is included; also in this case the changes are more progressive.
The non--rotating model has almost no WNL phase and does not enter the WC
stage. The model with rotation is very different, it  has an extended WNL phase
and then a remarkably large transition phase, where $^{14}$N is present with
a significant abundance, while $^{12}$C, $^{16}$O and $^{22}$Ne are also
there. Most of this transition phase is likely spent in the transition WN/WC phase,
but probably not the whole, so that some peculiar WN stars with products 
of He--burning are possible. Also some WC stars with an unusual 
occurrence of $^{14}$N are very likely. Globally, we see that the 
transition phase where both characteristics of H-- and He--burning phases
are present is longer for stars of lower masses (see Sect. 5).

\begin{figure}[t]
  \resizebox{\hsize}{!}{\includegraphics{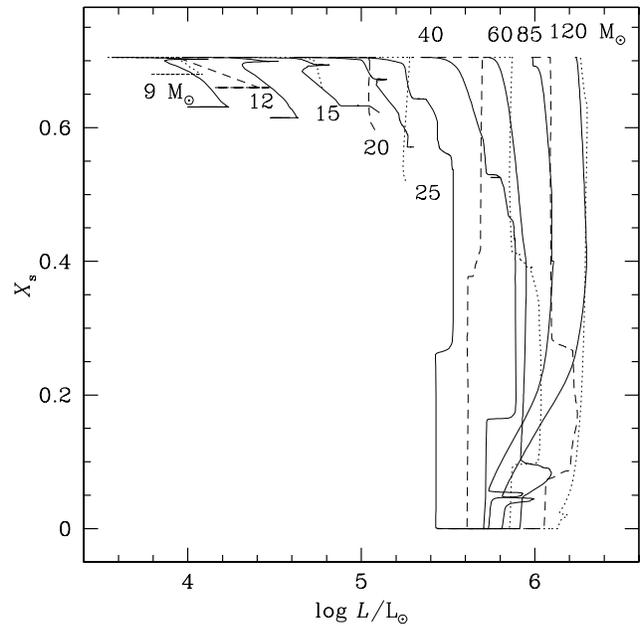}}
  \caption{Evolutionary tracks in the $X_{\rm s}$ versus log $L/L_\odot$ plane,
where $X_{\rm s}$ is the hydrogen mass fraction at the surface. The continuous lines
are for the rotating models, the dotted or dashed--lines are for the non--rotating models.}
  \label{xsl}
\end{figure}

Fig.~\ref{xsl} shows an interesting diagram for the study of
so--called ``slash stars'' of type Ofpe/WN, of the LBV and of the WN stars. It presents 
the evolution of the surface H--content as a function of the luminosity.
The tracks go downwards in this diagram; first there is an initial 
brightening due to MS evolution, then the luminosity is about constant, except 
for $X_{\mathrm{s}} \leq 0.20 $ if the mass loss is very high. Thus
for stars above or equal to 25 M$_{\odot}$, this diagram is sensitive to both
mass loss and mixing. Since WN subtypes are sensitive to the optical
thickness of the winds, while $X_{\mathrm{s}}$ is sensitive to both mixing
and mass loss, one  should try to define the domains of this
diagram where ''slash stars'', LBV, WN late and WN early are each located on
average. Ideally this should also be attempted for stars in different 
galaxies. For masses below 25 M$_{\odot}$, this diagram indicates
up to which stage mixing is proceeding for stars of different luminosities,
which is also a useful indication.

\begin{figure}[t]
  \resizebox{\hsize}{!}{\includegraphics{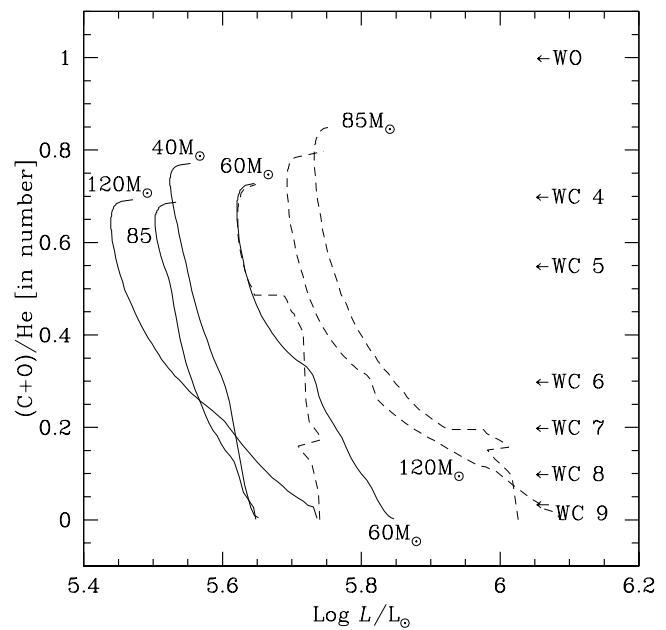}}
  \caption{Evolution of the ratios (C+O)/He as a function of the luminosity at the surface of non--rotating
  (dashed--lines) and rotating models (continuous lines) for various initial
  mass models. The correspondence
  between the (C+O)/He ratios and the different WC subtypes as given by Smith \& Maeder \cite{SmithM91}
  is indicated on the right of the figure.
}
  \label{cohe}
\end{figure}

The evolution of the (C+O)/He ratio as a function of luminosity is shown
in Fig.~\ref{cohe}, which is the key diagram for WC stars as shown by 
Smith \& Maeder (\cite{SmithM91}). The evolution for a given mass is upwards,
the luminosity decreases
as the stellar masses decreases. At the same time, as a result of mass loss
and mixing, the (C+O)/He ratio goes up, since there are more and more 
products of He--burning  at the stellar surface. 
{\emph{Rotation drastically modifies the tracks of WC stars in this
diagram, as well as the lifetimes}}.  The tracks of the models
of 85  and 120 M$_{\odot}$ are shifted to lower luminosities
by rotation. As there is a tight  mass--luminosity relation
for He, C, O stars (Maeder \cite{Maeder83}), this is the result of the smaller
mass at the entrance of the  WC phase for models with rotation. The smaller
mass is due mainly to the much longer WN phase of rotating models.
(As a minor remark, we see that the (C+O)/He ratio goes slightly higher in models without
rotation, this is due to the fact that the He-- and C--burning 
proceeds a bit farther in non rotating models that have a higher mass). 
For the models of 60 M$_{\odot}$, there is a small difference at the beginning of the WC phase
and this difference quickly disappears. The 40 M$_{\odot}$ models enter the 
WC phase only if rotation is significant.
In a previous work (Smith \& Maeder \cite{SmithM91}), we have shown that 
metallicity $Z$ strongly shifts the tracks in this diagram. Thus, we now
see that rotation is also producing great shifts. The location of WC stars
of different subtypes and composition should be attempted; it would be 
particularly worth considering WC stars of  galaxies of different $Z$.

\section{Conclusion}

The good agreement obtained here between the observed WR star populations
and the predictions of the rotating stellar models reinforces the general conclusion,
already obtained in previous works, namely that rotation is a key physical ingredient
of massive star evolution. Population synthesis for galaxies and starbursts
should also be based on stellar models with rotation. The same is true
for nucleosynthetic models.

Rotation, in addition to making the models more realistic,
opens new ways to explore very interesting questions.
Among them let we cite the possible relations between rotation and the ring nebulae observed 
around some WR stars, the relation between rotation and the LBV phenomenon, the rotation
of pulsars and the link to $\gamma$--ray bursts.


\begin{thebibliography}{}

\bibitem[1999]{Ang99}
Angulo, C, Arnould, M., Rayet, M. et al. 1999, Nucl. Phys. A, 656, 3

\bibitem[1989]{Co89}
Conti, P.S., \& Massey, P. 1989, ApJ, 337, 251

\bibitem[1995]{Cr95}
Crowther, P.A., Smith, L.J., \& Willis, A.J. 1995, A\&A, 304, 269

\bibitem[2000]{De00}
Dessart, L., Crowther, P.A., Hillier, J.D., Willis, A.J., Morris, P.W., \& van der
Hucht, K.A. 2000, MNRAS, 315, 407

\bibitem[1995]{Fl95}
Fliegner, J., \& Langer, N. 1995, IAU Symp. 163, ed. K. A van der Hucht \& P.M. Williams, (Dordrecht: Kluwer), 326

\bibitem[2003]{Fo03}
Foellmi, C., Moffat, A.F.J., \& Guerrero, M.A. 2003, MNRAS, 338, 1025

\bibitem[1986]{Fr86}
Friend, D.B., \& Abbott, D.C. 1986, A\&A, 311, 701

\bibitem[1999]{Ha99} 
Hamann, W.R., \& Koesterke, L. 1999. IAU Coll. 169, eds.  B. Wolf, O. Stahl, \& A.W. Fullerton,
(Berlin: Springer). Also Lecture Notes in Physics, 523, 1999, p. 239

\bibitem[1998]{he98}
Heger, A. 1998, phD, Max--Planck--Institut f\"ur Astrophysik, M\"unchen


\bibitem[2000]{He00}
Heger, A., \& Langer, N. 2000, ApJ, 544, 1016

 \bibitem[2000]{He20}
 Herrero, A., Puls, J., \& Villamariz, M.R. 2000, A\&A, 354, 193

\bibitem[2001]{Hu01}
van der Hucht, K.A. 2001, New Astronomy Reviews,  45, 135

\bibitem[1996]{IR96}
Iglesias, C.A., \& Rogers, F.J. 1996, ApJ, 231, 384


\bibitem[1988]{Ja88}
de Jager, C., Nieuwenhuijzen, H., \& van der Hucht, K.A. 1988,
        A\&AS, 72, 259

\bibitem[1991]{La91}
Langer, N. 1991, A\&A, 248, 531

\bibitem[1997]{La97}
Langer, N. 1997, The Eddington Limit in Rotating Massive Stars. In: Nota A., Lamers H.
(eds.) Luminous Blue Variables: Massive Stars 
in Transition. ASP Conf. Series, 120, p. 83

\bibitem[1971]{la71}
Lauterborn, D., Refsdal, S., \& Weigert, A. 1971, A\&A, 10, 97


\bibitem[1983]{Maeder83}
Maeder, A. 1983, A\&A, 120, 113

\bibitem[1987]{Maeder87}
Maeder, A. 1987, A\&A, 178, 159 

\bibitem[1991]{Maeder91}
Maeder, A. 1991, A\&A, 242, 93

\bibitem[1996]{Maeder96}
Maeder, A. 1996, in Wolf-Rayet stars in the framework 
of stellar evolution, ed.  J.M. Vreux et al. (Universit\'e de Li\`ege, 
Institut d'Astrophysique), 39


\bibitem[2002]{Ma02}
Maeder, A. 2002, A\&A, 392, 575

\bibitem[1989]{ma89}
Maeder, A., \& Meynet, G. 1989, A\&A, 210, 155


\bibitem[1994]{MaeMey94}
Maeder, A., \& Meynet, G. 1994, A\&A, 287, 803


\bibitem[2000a]{MMVI}
 Maeder, A., \& Meynet, G. 2000a, A\&A, 361, 159, (paper VI)

\bibitem[2000b]{araa}
Maeder, A., \& Meynet, G. 2000b, ARAA, 38, 143
 
\bibitem[2001]{MMVII}
 Maeder, A., \& Meynet, G. 2001, A\&A, 373, 555, (paper VII)

\bibitem[2002]{Mart02}
Martins, F., Schaerer, D., \& Hillier, D.J. 2002, A\&A, 382, 999

\bibitem[2003]{Mas03}
Massey, P. 2003, ARAA, in press 


\bibitem[2001]{Mas98}
Massey, P., \& Johnson, O. 1998, ApJ, 505, 793

\bibitem[2001]{Mas01}
Massey, P., \& Duffy, A.S. 2001, ApJ, 550, 713


\bibitem[2002]{Mas02}
Massey, P., \& Holmes, S. 2002, ApJ, 580, L35


\bibitem[2000]{Me00}
Meynet, G. 2000. In Massive Stellar Clusters, ASP Conf. Ser. 211,, eds A. Lan\c con \& C. M. Boily, p. 105

\bibitem[1997]{MMI}
Meynet, G., \& Maeder, A. 1997, A\&A, 321, 465, (paper I)

\bibitem[2000]{MMV}
Meynet, G., \& Maeder, A. 2000, A\&A, 361, 101, (paper V)

\bibitem[2002]{MMVIII}
Meynet, G., \& Maeder, A. 2002, A\&A, 390, 561, (paper VIII)


\bibitem[1994]{Mey94}
 Meynet, G., Maeder, A., Schaller,
 G., Schaerer, D., \& Charbonnel, C. 1994, A\&A Suppl., 103, 97

\bibitem[2000]{NuLa00}
Nugis, T., \& Lamers, H.J.G.L.M. 2000, A\&A, 360, 227

\bibitem[1970]{Sa70}
Sackman, I.J., \& Anand, S.P.S. 1970, ApJ, 162, 105

\bibitem[1996]{Sc96}
Schaerer, D. 1996, ApJ, 467, L17


\bibitem[1992]{Sc92}
Schaerer, D., \& Maeder, A. 1992, A\&A, 263, 129

\bibitem[1992]{Sch92}
Schaller, G., Schaerer, D., Meynet, G., \& Maeder, A.
      1992, A\&A Suppl., 96, 269


\bibitem[2002]{Sch02}
Schild, H., Crowther, P.A., Abbott, J.B., \& Schmutz, W. 2003, A\&A, 397, 859

\bibitem[2002]{spruit02}
Spruit, H. 2002, A\&A, 381, 923

  
\bibitem[1991]{SmithM91}
Smith, L.F., \& Maeder, A. 1991, A\&A, 241, 77

\bibitem[1997]{TaZa97}
Talon, S., \& Zahn, J.P. 1997, A\&A, 317, 749   

\bibitem[2000]{Vink00}
Vink, J.S., de Koter, A., \& Lamers, H.J.G.L.M. 2000, A\&A, 362, 295

\bibitem[2001]{Vink01}
Vink, J.S., de Koter, A., \& Lamers, H.J.G.L.M. 2001, A\&A, 369, 574

\bibitem[1999]{Wi99} 
Willis, A.J. 1999, IAU Symp. 193, K.A. van der Hucht, G. Koenigsberger, and P.R.J.
Eenens eds., p. 1


\bibitem[2003]{Woo03}
Woosley, S. 2003, IAU Symp. 215, Eds. A. Maeder \& P. Eenens,
in press.

\end{thebibliography}
\end{document}